\documentclass[twocolumn]{autart}
\usepackage{amssymb}
\usepackage{amsmath}
\usepackage{graphicx}

\input{tcilatex}

\begin{document}

\begin{frontmatter}

\title{Linearly Constrained Kalman Filter For Linear Discrete State-Space Models\thanksref{footnoteinfo}} 

\thanks[footnoteinfo]{This paper was not presented at any IFAC
meeting. This work has been partially supported by
the DGA/MRIS (2015.60.0090.00.470.75.01). Corresponding author M.~E.~Chaumette. Tel. +33561338925.}

\author[Isae]{Eric Chaumette}\ead{eric.chaumette@isae.fr},    
\author[Isae]{Francois Vincent}\ead{francois.vincent@isae.fr}  

\address[Isae]{Universit\'{e} de Toulouse/ISAE-Supa\'{e}ro, 10 avenue Edouard Belin, 31400 Toulouse}  


\begin{abstract}                          
For linear discrete state-space (LDSS) models, under certain conditions, the
linear least mean squares filter estimate has a convenient recursive
predictor/corrector format, aka the Kalman filter (KF). The aim of the paper
is to introduce the general form of the linearly constrained KF (LCKF) for
LDSS models, which encompasses the linearly constrained minimum variance
estimator (LCMVE). Thus the LCKF opens access to the abundant litterature on
LCMVE in the deterministic framework which can be transposed to the
stochastic framework. Therefore, among other things, the LCKF may provide
alternative solutions to $H_{\infty }$ filter and unbiased finite impulse
response filter to robustify the KF, which performance are sensible to
misspecified noise or uncertainties in the system matrices.
\end{abstract}

\end{frontmatter}

\section{Introduction}

We consider the general class of linear discrete state-space (LDSS) models
represented with the state and measurement equations, respectively,
\begin{subequations}
\begin{eqnarray}
\mathbf{x}_{k} &=&\mathbf{F}_{k-1}\mathbf{x}_{k-1}+\mathbf{w}_{k-1}
\label{DLS - recursion - xk} \\
\mathbf{y}_{k} &=&\mathbf{H}_{k}\mathbf{x}_{k}+\mathbf{v}_{k}
\label{DLS - recursion - yk}
\end{eqnarray}%
where the time index $k\geq 1$, $\mathbf{x}_{k}$ is the $P_{k}$-dimensional
state vector, $\mathbf{y}_{k}$ is the $N_{k}$-dimensional measurement
vector. The state and measurement noise sequences $\left\{ \mathbf{w}%
_{k}\right\} $ and $\left\{ \mathbf{v}_{k}\right\} $, as well as the initial
state $\mathbf{x}_{0}$ are random vectors with arbitrary distributions. The
noise sequences $\left\{ \mathbf{w}_{k}\right\} $ and $\left\{ \mathbf{v}%
_{k}\right\} $ have zero-mean values\footnote{%
This assumption is equivalent to the assumption of nonzero but known noises
mean values \cite[\S 3.2.4]{Kailath - Sayed - Hassibi}.} and the initial
state $\mathbf{x}_{0}$ has a finite known mean value. The system matrices $%
\left\{ \mathbf{F}_{k},\mathbf{H}_{k}\right\} $ and the covariance and
cross-covariance matrices of $\left\{ \mathbf{w}_{k},\mathbf{v}_{k},\mathbf{x%
}_{0}\right\} $ contain elements with finite modulus and are either known or
specified according to known parametric models. The objective is to estimate
$\mathbf{x}_{k}$ based on the measurements and our knowledge of the model
dynamics. If the estimate of $\mathbf{x}_{k}$ is based on measurements up to
and including time $l$, we denote the estimator as $\widehat{\mathbf{x}}%
_{k|l}\triangleq \widehat{\mathbf{x}}_{k|l}\left( \mathbf{y}_{1},\ldots ,%
\mathbf{y}_{l}\right) $ and we use the term estimator to refer to the class
of algorithms that includes filtering, prediction, and smoothing. A filter
estimates $\mathbf{x}_{k}$ based on measurements up to and including time $k$%
. A predictor estimates $\mathbf{x}_{k}$ based on measurements prior to time
$k$. A smoother estimates $\mathbf{x}_{k}$ based on measurements prior to
time $k$, at time $k$, and later than time $k$. Since the seminal paper of
Kalman \cite{Kalman}, it is known that, provided that the system matrices $%
\left\{ \mathbf{F}_{k},\mathbf{H}_{k}\right\} $ and the covariance and
cross-covariance matrices of $\left\{ \mathbf{w}_{k},\mathbf{v}_{k},\mathbf{x%
}_{0}\right\} $ are known, if $\left\{ \mathbf{w}_{k},\mathbf{v}_{k},\mathbf{%
x}_{0}\right\} $ verify certain uncorrelation conditions \cite[(18)]%
{Chaumette - TAC} and are Gaussian, the minimum variance or minimum mean
squared error (MSE) filter estimate for LDSS models has a convenient
recursive predictor/corrector format\footnote{%
The superscript $^{b}$ is used to remind the reader that the value under
consideration is the "best" one according to a criterion previously defined.}%
, $\forall k\geq 2$:
\end{subequations}
\begin{equation}
\widehat{\mathbf{x}}_{k|k}^{b}=\mathbf{F}_{k-1}\widehat{\mathbf{x}}%
_{k-1|k-1}^{b}+\mathbf{K}_{k}^{b}\left( \mathbf{y}_{k}-\mathbf{H}_{k}\mathbf{%
F}_{k-1}\widehat{\mathbf{x}}_{k-1|k-1}^{b}\right) ,
\label{KF - pred/cor format}
\end{equation}%
so-called the Kalman filter (KF) \cite{Kalman}. Even if the noise is
non-Gaussian, the KF is the linear least mean squares (LLMS) filter (LLMSF)
estimate \cite{Wiener}. As the computation of the KF depends on prior
information on the first and second order statistics of the initial state $%
\mathbf{x}_{0}$ \cite{Kailath - Sayed - Hassibi}\cite{Simon}\cite{Gibbs},
the KF can be looked upon as an "initial state first and second order
statistics" matched filter \cite{Chaumette - TAC}.\ However in numerous
applications first and second order statistics of $\mathbf{x}_{0}$ may be
unknown. A commonly used solution to circumvent this lack of prior
information on $\mathbf{x}_{0}$ is the Fisher initialization \cite{Catlin}%
\cite[\S II]{Chen}. The Fisher initialization consists in initializing the
KF recursion at time $k=1$ with the best linear unbiased estimator (BLUE) of
$\mathbf{x}_{1}$ associated to the measurement model (\ref{DLS - recursion -
yk}), where $\mathbf{x}_{1}$ is regarded as a deterministic unknown
parameter vector. In the deterministic framework, the BLUE of $\mathbf{x}_{1}
$ is also known as the linear minimum variance distortionless response
(LMVDR) estimator of $\mathbf{x}_{1}$ \cite[\S 6]{Van Trees Part IV}\cite[\S %
5.6]{Schreier - Scharf}\cite{Vorobyov} and coincides with the weighted least
squares estimator (WLSE) of $\mathbf{x}_{1}$. If $\mathbf{H}_{1}$ is full
rank and the covariance matrix of $\mathbf{v}_{1}$ ($\mathbf{C}_{\mathbf{v}%
_{1}}$) is invertible, the Fisher initialization yields:
\begin{equation}
\widehat{\mathbf{x}}_{1|1}=\mathbf{P}_{1|1}\mathbf{H}_{1}^{H}\mathbf{C}_{%
\mathbf{v}_{1}}^{-1}\mathbf{y}_{1},~\mathbf{P}_{1|1}=\left( \mathbf{H}%
_{1}^{H}\mathbf{C}_{\mathbf{v}_{1}}^{-1}\mathbf{H}_{1}\right) ^{-1}.
\label{WLSE x1}
\end{equation}%
Actually, under mild regularity conditions on the noises covariance
matrices, the Fisher initialization (\ref{WLSE x1}) yields the stochastic
LMVDR filter (LMVDRF), which shares the same recursion as the KF, except at
time $k=1$ \cite{Chaumette - TAC}\cite{Chaumette - SPL}. Although the LMVDRF
is sub-optimal in MSE sense and is an upper bound on the performance of the
KF, it is an infinite impulse response distortionless filter which
performance is robust to an unknown initial state. However since the LMVDRF
shares the same recursion as the KF, it also shares the same sensitivity to
misspecified covariance matrices \cite[\S 10]{Simon}\cite{Heffes}\cite{Zhang}%
\cite{Hsiao} or uncertainties in the system matrices \cite{Pena - Guttman}%
\cite{Theodor - Shaked}\cite{Xie94}\cite{Kosanam - Simon}. This sensibility
of the performance achievable by the LMVDR estimator to misspecifications or
uncertainties is also well documented in deterministic parameters estimation
\cite[\S\ 6.7]{Van Trees Part IV}\cite{Vorobyov}. For instance, in array
processing, the performance of MVDR beamformers are not particularly robust
in the presence of various types of differences between the model and the
actual environment (array perturbation, direction of arrival mismatch,
inaccurate estimation of $\mathbf{C}_{\mathbf{v}_{k}}$, ...) \cite[\S\ 6.6]%
{Van Trees Part IV}\cite{Vorobyov}. Thus linearly constrained minimum
variance (LCMV) beamformers have been developed in which additional linear
constraints are imposed to make the MVDR beamformer more robust \cite[\S\ 6.7%
]{Van Trees Part IV}\cite{Vorobyov}.

The aim of the paper is to introduce the general form of the linearly
constrained KF (LCKF) for LDSS models. Among other things, the LCKF can be
used to robustify the KF, where robustness is understood as an ability to
achieve high performance in the situations with imperfect, incomplete, or
erroneous knowledge about the system under consideration and its
environment. So far, in many applications where the statistical properties
of state and measurement noises are not accurately known, it has been common
practice to use a $H_{\infty }$ filter \cite[\S 10]{Simon}\cite{Xie04}\cite%
{Banavar}\cite{Shen}, also called the minimax filter, since it does not make
any assumptions about the noise, and it minimizes the worst-case estimation
error. Lately another possible way to robustify the KF to the presence of
noises mismodeling via unbiased finite impulse response (UFIR) \cite{Shmaliy}%
, p-shift FIR \cite{Shmaliy 2011}\cite[\S 11]{Simon - Shmaliy} or minimum
variance UFIR \cite{Zhao} filters, has been introduced. These algorithms
have the same predictor/corrector format as the KF, often ignore initial
estimations errors and the statistics of the noise, and become virtually
optimal as the length of the FIR window increases. Therefore, since the LCMV
estimator (LCMVE) is a special case of the LCKF, the use of LCKF opens
access to the abundant literature on LCMVE in the deterministic framework
\cite{Vorobyov} which can be transposed to the stochastic framework in order
to provide alternative solutions to $H_{\infty }$ filter and UFIR filter to
robustify the KF. As an example, we show how linear constraints can be used
to robustify the KF in the presence of parametric modelling errors in the
system matrices $\left\{ \mathbf{F}_{k},\mathbf{H}_{k}\right\} $. However,
the disadvantage of using multiple linear constraints is that additional
degrees of freedom are used by the LCKF in order to satisfy these
constraints which increases the minimum MSE achieved. Last, it is noteworthy
that linear constraints can be taken into account in any existing
generalizations of the KF \cite[\S 7]{Simon}, whether to deal with
correlated state and measurement noise, colored state noise, colored
measurement noise, for filtering with fading memory, to incorporate state
constraints, for prediction, for smoothing, ....

The rest of the paper is organized as follows. Notations and signal model
(joint proper complex) are introduced in Section II. In Section III, for
sake of clarity, we give the main points of background knowledge on linear
filters (including LLMSF and LMVDRF) required to discuss the filtering%
\textit{\ }equations in the next Section. In section IV, we derive the
general form of the LCKF for LDSS models and provide some analysis on the
various forms of the LCKF recursion which depends on linear constraints
combination. In section V, we show that the LCMVE in deterministic
parameters estimation is a special case of the LCKF, which opens access to
the abundant literature on LCMVE. Last, an example of the possible
transposition of the LCMVE's literature to the stochastic framework is given
in Section VI.

\section{Notations and signal model}

The notational convention adopted is as follows: we shall use italic, small
boldface and capital boldface letters to denote respectively scalars, column
vectors and matrices. $\mathcal{M}_{%
\mathbb{C}
}\left( N,P\right) $ denotes the vector space of complex matrices with $N$
rows and $P$ columns. The scalar/matrix/vector transpose conjugate is
indicated by the superscript $^{H}$. $\mathbf{I}$ is the identity matrix. $%
\left[ \mathbf{A}~\mathbf{B}\right] $ and $\QTATOPD[ ] {\mathbf{A}}{\mathbf{B%
}}$ denote the matrix resulting from the horizontal and the vertical
concatenation of $\mathbf{A}$ and $\mathbf{B}$,\textbf{\ }respectively. The
matrix resulting from the vertical concatenation $k$ matrices $\mathbf{A}_{1}
$, ..., $\mathbf{A}_{k}$ of same column number is denoted $\overline{\mathbf{%
A}}_{k}$. $E\left[ \cdot \right] $ denotes the expectation operator. If $%
\mathbf{x}$ and $\mathbf{y}$ are two complex random vectors: a) $\mathbf{C}_{%
\mathbf{x}}$, $\mathbf{C}_{\mathbf{y}}$ and $\mathbf{C}_{\mathbf{x,y}}$ are
respectively the covariance matrices of $\mathbf{x}$, of $\mathbf{y}$ and
the cross-covariance matrix of $\mathbf{x}$ and $\mathbf{y}$; b) if $\mathbf{%
C}_{\mathbf{y}}$ is invertible, then $\mathbf{C}_{\mathbf{x|y}}\triangleq
\mathbf{C}_{\mathbf{x}}-\mathbf{C}_{\mathbf{x},\mathbf{y}}\mathbf{C}_{%
\mathbf{y}}^{-1}\mathbf{C}_{\mathbf{x},\mathbf{y}}^{H}$\footnote{%
If $\mathbf{x}$ and $\mathbf{y}$ are (proper) normal complex random vectors,
then $\mathbf{C}_{\mathbf{x|y}}$ is the covariance matrix of $\mathbf{x}$
given $\mathbf{y}$.}.\newline
As in \cite[\S 3]{Kailath - Sayed - Hassibi} and \cite[\S 5.1]{Schreier -
Scharf}, we adopt a joint proper (proper and cross-proper) complex signals
assumption for the set of vector $\left( \mathbf{x}_{0},\left\{ \mathbf{w}%
_{k}\right\} ,\left\{ \mathbf{v}_{k}\right\} \right) $ which allows to
resort to standard estimation in the MSE sense defined on the Hilbert space
of complex random variables with finite second-order moment. A proper
complex random variable is uncorrelated with its complex conjugate \cite%
{Schreier - Scharf}, and a zero mean proper complex random vector is said to
be second-order circular \cite[\S 3.2.5]{Kailath - Sayed - Hassibi}.
Moreover, any result derived with joint proper complex random vectors are
valid for real random vectors provided that one substitutes the
matrix/vector transpose conjugate for the matrix/vector transpose \cite[\S %
3.2.5]{Kailath - Sayed - Hassibi}\cite[\S 5.4.1]{Schreier - Scharf}.

\subsection{Equivalent linear observation model}

Here, $\mathbf{F}_{k-1}\in \mathcal{M}_{%
\mathbb{C}
}\left( P_{k},P_{k-1}\right) $ and $\mathbf{H}_{k}\in \mathcal{M}_{%
\mathbb{C}
}\left( N_{k},P_{k}\right) $. First, as (\ref{DLS - recursion - xk}) can be
rewritten as, $\forall k\geq 2$:
\begin{equation*}
\mathbf{x}_{k}=\mathbf{B}_{k,1}\mathbf{x}_{1}+\tsum_{l=1}^{k-1}\mathbf{B}%
_{k,l+1}\mathbf{w}_{l},\mathbf{B}_{k,l}=\left\vert
\begin{array}{r}
\mathbf{F}_{k-1}\mathbf{F}_{k-2}...\mathbf{F}_{l},k>l \\
\mathbf{I\qquad },k=l \\
\mathbf{0\qquad },k<l%
\end{array}%
\right.
\end{equation*}%
an equivalent form of (\ref{DLS - recursion - yk}) is:%
\begin{multline}
\mathbf{y}_{k}=\mathbf{A}_{k}\mathbf{x}_{1}+\mathbf{n}_{k},\mathbf{~A}_{k}=%
\mathbf{H}_{k}\mathbf{B}_{k,1}, \\
\left\vert
\begin{array}{l}
\mathbf{n}_{1}=\mathbf{v}_{1} \\
\mathbf{n}_{k}=\tsum_{l=1}^{k-1}\mathbf{H}_{k}\mathbf{B}_{k,l+1}\mathbf{w}%
_{l}+\mathbf{v}_{k},\mathbf{~}k\geq 2%
\end{array}%
\right. .
\end{multline}%
Second, (\ref{DLS - recursion - yk}) can be extended on a horizon of $k$
points from the first observation as:
\begin{equation}
\overline{\mathbf{y}}_{k}=\left(
\begin{array}{c}
\mathbf{y}_{1} \\
\vdots \\
\mathbf{y}_{k}%
\end{array}%
\right) =\left[
\begin{array}{c}
\mathbf{A}_{1} \\
\vdots \\
\mathbf{A}_{k}%
\end{array}%
\right] \mathbf{x}_{1}+\left(
\begin{array}{c}
\mathbf{n}_{1} \\
\vdots \\
\mathbf{n}_{k}%
\end{array}%
\right) =\overline{\mathbf{A}}_{k}\mathbf{x}_{1}+\overline{\mathbf{n}}_{k},
\end{equation}%
$\overline{\mathbf{y}}_{k},\overline{\mathbf{n}}_{k}\in \mathcal{M}_{%
\mathbb{C}
}\left( \mathcal{N}_{k},1\right) $, $\overline{\mathbf{A}}_{k}\in \mathcal{M}%
_{%
\mathbb{C}
}\left( \mathcal{N}_{k},P_{k}\right) $, $\mathcal{N}_{k}=\tsum%
\nolimits_{l=1}^{k}N_{l}$.

\section{Background on linear filters\label{S: Background on linear
estimators}}

\subsection{Linear least-mean-squares estimator (LLMSE)}

Let us consider two joint zero mean proper complex random vectors $\mathbf{x}
$ and $\mathbf{y}$. The error between the signal $\mathbf{x}$ and the linear
estimator $\widehat{\mathbf{x}}\triangleq \widehat{\mathbf{x}}\left( \mathbf{%
y}\right) =\mathbf{Ky}$, $\mathbf{K}\in \mathcal{M}_{%
\mathbb{C}
}\left( \dim \left( \mathbf{x}\right) ,\dim \left( \mathbf{y}\right) \right)
$, is $\mathbf{e}\triangleq \mathbf{e}\left( \mathbf{y},\mathbf{x}\right) =%
\mathbf{\widehat{\mathbf{x}}\left( \mathbf{y}\right) -x}$ and the error
covariance matrix is:
\begin{equation}
\mathbf{P}\left( \mathbf{K}\right) =E\left[ \mathbf{ee}^{H}\right] =E\left[
\left( \mathbf{\widehat{\mathbf{x}}\left( \mathbf{y}\right) -x}\right)
\left( \mathbf{\widehat{\mathbf{x}}\left( \mathbf{y}\right) -x}\right) ^{H}%
\right] .  \label{MSE - Def}
\end{equation}%
If $\mathbf{C}_{\mathbf{y}}$ is invertible, then (\ref{MSE - Def}) can be
rewritten as \cite{Wiener}\cite[p121]{Schreier - Scharf}:%
\begin{equation}
\mathbf{P}\left( \mathbf{K}\right) =\mathbf{C}_{\mathbf{x|y}}+\left( \mathbf{%
K}-\mathbf{C}_{\mathbf{x},\mathbf{y}}\mathbf{C}_{\mathbf{y}}^{-1}\right)
\mathbf{C}_{\mathbf{y}}\left( \mathbf{K}-\mathbf{C}_{\mathbf{x},\mathbf{y}}%
\mathbf{C}_{\mathbf{y}}^{-1}\right) ^{H},  \label{P(K) - LLMSE - Factorized}
\end{equation}%
yielding:
\begin{subequations}
\begin{multline}
\mathbf{K}^{b}=\arg \underset{\mathbf{K}}{\min }\left\{ \mathbf{P}\left(
\mathbf{K}\right) \right\} =\mathbf{C}_{\mathbf{x},\mathbf{y}}\mathbf{C}_{%
\mathbf{y}}^{-1},  \label{LLMSE- min(P(K))} \\
\mathbf{P}\left( \mathbf{K}^{b}\right) =\mathbf{C}_{\mathbf{x|y}},
\end{multline}%
\begin{equation}
\widehat{\mathbf{x}}^{b}=\mathbf{C}_{\mathbf{x},\mathbf{y}}\mathbf{C}_{%
\mathbf{y}}^{-1}\mathbf{y}.  \label{LLMSE}
\end{equation}

\subsection{Linear least-mean-squares filter (LLMSF)}

Therefore, the LLMSF of $\mathbf{x}_{k}$ based on measurements up to and
including time $k$, $k\geq 2$, is simply \cite{Kalman}\cite{Wiener}:
\end{subequations}
\begin{subequations}
\begin{equation}
\widehat{\mathbf{x}}_{k|k}^{b}=\left[ \mathbf{G}_{k-1}^{b}\ \mathbf{K}%
_{k}^{b}\right] \overline{\mathbf{y}}_{k}~|~\left[ \mathbf{G}_{k-1}^{b}\
\mathbf{K}_{k}^{b}\right] \mathbf{C}_{\overline{\mathbf{y}}_{k}}=\mathbf{C}_{%
\mathbf{x}_{k},\overline{\mathbf{y}}_{k}},  \label{LLMSF}
\end{equation}%
where $\mathbf{G}_{k-1}^{b}\in \mathcal{M}_{%
\mathbb{C}
}\left( P_{k},\mathcal{N}_{k-1}\right) $ and $\mathbf{K}_{k}^{b}\in \mathcal{%
M}_{%
\mathbb{C}
}\left( P_{k},N_{k}\right) $. A few lines of algebra allows to rewrite (\ref%
{LLMSF}) as \cite{Chaumette - SPL}:
\begin{equation}
\widehat{\mathbf{x}}_{k|k}^{b}=\widehat{\mathbf{x}}_{k|k-1}^{b}+\mathbf{K}%
_{k}^{b}\left( \mathbf{y}_{k}-\widehat{\mathbf{y}}_{k|k-1}^{b}\right) ,
\label{LLMSF - pred/cor (1)}
\end{equation}%
which is the general form of the so-called predictor/corrector format of
LLMSF ($\widehat{\mathbf{x}}_{k|k-1}^{b}$ is also known as the a priori
estimate of $\mathbf{x}_{k}$). Moreover, (\ref{LLMSF - pred/cor (1)}) can be
recasted as \cite{Chaumette - TAC}\cite{Chaumette - SPL}:%
\begin{multline}
\widehat{\mathbf{x}}_{k|k}^{b}=\left( \mathbf{I}-\mathbf{K}_{k}^{b}\mathbf{H}%
_{k}\right) \mathbf{F}_{k-1}\widehat{\mathbf{x}}_{k-1|k-1}^{b}+\mathbf{K}%
_{k}^{b}\mathbf{y}_{k}  \label{LLMSF - pred/cor (2)} \\
+\left( \mathbf{I}-\mathbf{K}_{k}^{b}\mathbf{H}_{k}\right) \widehat{\mathbf{w%
}}_{k-1|k-1}^{b}-\mathbf{K}_{k}^{b}\widehat{\mathbf{v}}_{k|k-1}^{b},~k\geq 2,
\end{multline}%
where $\widehat{\mathbf{v}}_{k|k-1}^{b}=\mathbf{C}_{\mathbf{v}_{k},\overline{%
\mathbf{y}}_{k-1}}\mathbf{C}_{\overline{\mathbf{y}}_{k-1}}^{-1}\overline{%
\mathbf{y}}_{k-1}$ and $\widehat{\mathbf{w}}_{k-1|k-1}^{b}=\mathbf{C}_{%
\mathbf{w}_{k-1},\overline{\mathbf{y}}_{k-1}}\mathbf{C}_{\overline{\mathbf{y}%
}_{k-1}}^{-1}\overline{\mathbf{y}}_{k-1}$. Thus, the general assumptions
required to obtain the recursive form (\ref{KF - pred/cor format}) of the
LLMSF (\ref{LLMSF - pred/cor (2)}), aka the KF, are:
\end{subequations}
\begin{subequations}
\begin{equation}
\widehat{\mathbf{w}}_{k-1|k-1}^{b}=\mathbf{0},~\ \widehat{\mathbf{v}}%
_{k|k-1}^{b}=\mathbf{0},~\forall \overline{\mathbf{y}}_{k-1},~k\geq 2,
\end{equation}%
that is:%
\begin{equation}
\mathbf{C}_{\mathbf{w}_{k-1},\overline{\mathbf{y}}_{k-1}}=\mathbf{0,~C}_{%
\mathbf{v}_{k},\overline{\mathbf{y}}_{k-1}}=\mathbf{0,}~k\geq 2.
\label{LLMSF - recursive pred/cor cond}
\end{equation}%
Another noteworthy point is that under the general assumptions (\ref{LLMSF -
recursive pred/cor cond}), the MSE of any linear filter $\widehat{\mathbf{x}}%
_{k|k}=\left[ \mathbf{G}_{k-1}\ \mathbf{K}_{k}\right] \overline{\mathbf{y}}%
_{k}$, $\mathbf{G}_{k-1}\in \mathcal{M}_{%
\mathbb{C}
}\left( P_{k},\mathcal{N}_{k-1}\right) $ and $\mathbf{K}_{k}\in \mathcal{M}_{%
\mathbb{C}
}\left( P_{k},N_{k}\right) $, that is:
\end{subequations}
\begin{equation}
\mathbf{P}_{k|k}\left( \mathbf{G}_{k-1},\mathbf{K}_{k}\right) =E\left[
\left( \widehat{\mathbf{x}}_{k|k}-\mathbf{x}_{k}\right) \left( \widehat{%
\mathbf{x}}_{k|k}-\mathbf{x}_{k}\right) \right] ,  \label{MSE LF}
\end{equation}%
breaks down into \cite{Chaumette - SPL}:
\begin{subequations}
\begin{multline}
\mathbf{P}_{k|k}\left( \mathbf{G}_{k-1},\mathbf{K}_{k}\right) =\mathbf{Q}%
_{k-1}\left( \mathbf{G}_{k-1},\mathbf{K}_{k}\right)
\label{MSE factorization} \\
+\left( \mathbf{I}-\mathbf{K}_{k}\mathbf{H}_{k}\right) \left(
\begin{array}{l}
\mathbf{C}_{\mathbf{w}_{k-1}}+\mathbf{F}_{k-1}\mathbf{C}_{\mathbf{w}_{k-1},%
\mathbf{x}_{k-1}}^{H} \\
+\mathbf{C}_{\mathbf{w}_{k-1},\mathbf{x}_{k-1}}\mathbf{F}_{k-1}^{H}%
\end{array}%
\right) \left( \mathbf{I}-\mathbf{K}_{k}\mathbf{H}_{k}\right) ^{H} \\
-\left( \mathbf{I}-\mathbf{K}_{k}\mathbf{H}_{k}\right) \mathbf{C}_{\mathbf{x}%
_{k},\mathbf{v}_{k}}\mathbf{K}_{k}^{H}-\mathbf{K}_{k}\mathbf{C}_{\mathbf{x}%
_{k},\mathbf{v}_{k}}^{H}\left( \mathbf{I}-\mathbf{K}_{k}\mathbf{H}%
_{k}\right) ^{H} \\
+\mathbf{K}_{k}\mathbf{C}_{\mathbf{v}_{k}}\mathbf{K}_{k}^{H},
\end{multline}%
where:%
\begin{multline}
\mathbf{Q}_{k-1}\left( \mathbf{G}_{k-1},\mathbf{K}_{k}\right) =
\label{MSE factorization - Q} \\
E\left[
\begin{array}{c}
\left( \mathbf{G}_{k-1}\overline{\mathbf{y}}_{k-1}-\left( \mathbf{I}-\mathbf{%
K}_{k}\mathbf{H}_{k}\right) \mathbf{F}_{k-1}\mathbf{x}_{k-1}\right) \times
\\
\left( \mathbf{G}_{k-1}\overline{\mathbf{y}}_{k-1}-\left( \mathbf{I}-\mathbf{%
K}_{k}\mathbf{H}_{k}\right) \mathbf{F}_{k-1}\mathbf{x}_{k-1}\right) ^{H}%
\end{array}%
\right] ,
\end{multline}%
which is a key result in order to derive the general form of the KF and
LMVDR filter recursion (without extension of the state and measurement
equations). Indeed, from (\ref{MSE factorization}), it is obvious that:
\end{subequations}
\begin{subequations}
\begin{equation}
\mathbf{G}_{k-1}^{b}=\arg \underset{\mathbf{G}_{k-1}}{\min }\left\{ \mathbf{Q%
}_{k-1}\left( \mathbf{G}_{k-1},\mathbf{K}_{k}\right) \right\} ,
\end{equation}%
that is (\ref{LLMSE}):
\begin{multline}
\mathbf{G}_{k-1}^{b}\overline{\mathbf{y}}_{k-1}=\mathbf{C}_{\left( \mathbf{I}%
-\mathbf{K}_{k}\mathbf{H}_{k}\right) \mathbf{F}_{k-1}\mathbf{x}_{k-1},%
\overline{\mathbf{y}}_{k-1}}\mathbf{C}_{\overline{\mathbf{y}}_{k-1}}^{-1}%
\overline{\mathbf{y}}_{k-1} \\
=\left( \mathbf{I}-\mathbf{K}_{k}\mathbf{H}_{k}\right) \mathbf{F}_{k-1}%
\widehat{\mathbf{x}}_{k-1|k-1}^{b},
\end{multline}%
leading to the general form of the Joseph stabilized version of the
covariance measurement update equation \cite{Chaumette - TAC}\cite{Chaumette
- SPL}:
\end{subequations}
\begin{multline}
\mathbf{P}_{k|k}\left( \mathbf{G}_{k-1}^{b},\mathbf{K}_{k}\right) =\left(
\mathbf{I}-\mathbf{K}_{k}\mathbf{H}_{k}\right) \mathbf{P}_{k|k-1}^{b}\left(
\mathbf{I}-\mathbf{K}_{k}\mathbf{W}_{k}\right) ^{H}  \label{Joseph} \\
-\left( \mathbf{I}-\mathbf{K}_{k}\mathbf{H}_{k}\right) \mathbf{C}_{\mathbf{x}%
_{k},\mathbf{v}_{k}}\mathbf{K}_{k}^{H}-\mathbf{K}_{k}\mathbf{C}_{\mathbf{x}%
_{k},\mathbf{v}_{k}}^{H}\left( \mathbf{I}-\mathbf{K}_{k}\mathbf{H}%
_{k}\right) ^{H} \\
+\mathbf{K}_{k}\mathbf{C}_{\mathbf{v}_{k}}\mathbf{K}_{k}^{H}
\end{multline}%
where:
\begin{eqnarray*}
\mathbf{P}_{k|k-1}^{b} &=&\mathbf{F}_{k-1}\mathbf{P}_{k-1|k-1}^{b}\mathbf{F}%
_{k-1}^{H}+\mathbf{C}_{\mathbf{w}_{k-1}} \\
&&+\mathbf{F}_{k-1}\mathbf{C}_{\mathbf{w}_{k-1},\mathbf{x}_{k-1}}^{H}+%
\mathbf{C}_{\mathbf{w}_{k-1},\mathbf{x}_{k-1}}\mathbf{F}_{k-1}^{H} \\
&=&E\left[ \left( \widehat{\mathbf{x}}_{k|k-1}^{b}-\mathbf{x}_{k}\right)
\left( \widehat{\mathbf{x}}_{k|k-1}^{b}-\mathbf{x}_{k}\right) ^{H}\right] .
\end{eqnarray*}%
The solution of the minimization of (\ref{Joseph}) is well known, since (\ref%
{Joseph}) can be reformulated as \cite{Kailath - Sayed - Hassibi}\cite{Simon}%
\cite{Gibbs}:
\begin{subequations}
\begin{equation}
\mathbf{P}_{k|k}\left( \mathbf{G}_{k-1}^{b},\mathbf{K}_{k}\right) =E\left[
\begin{array}{c}
\left( \mathbf{K}_{k}\mathbf{\varepsilon }_{k}-\left( \mathbf{x}_{k}-%
\widehat{\mathbf{x}}_{k|k-1}^{b}\right) \right) \times  \\
\left( \mathbf{K}_{k}\mathbf{\varepsilon }_{k}-\left( \mathbf{x}_{k}-%
\widehat{\mathbf{x}}_{k|k-1}^{b}\right) \right) ^{H}%
\end{array}%
\right] ,  \label{Pk|k - innovation}
\end{equation}%
where:
\begin{multline}
\mathbf{\varepsilon }_{k}=\mathbf{H}_{k}\left( \mathbf{x}_{k}-\widehat{%
\mathbf{x}}_{k|k-1}^{b}\right) +\mathbf{v}_{k}=\mathbf{y}_{k}-\mathbf{H}_{k}%
\widehat{\mathbf{x}}_{k|k-1}^{b}  \label{innovations vector} \\
=\mathbf{y}_{k}-\widehat{\mathbf{y}}_{k|k-1}^{b},
\end{multline}%
is the innovations vector. Thus, according to (\ref{LLMSE- min(P(K))}), $%
\mathbf{K}_{k}^{b}$ is computed according to the following recursion for $%
k\geq 2$:
\end{subequations}
\begin{subequations}
\begin{multline}
\mathbf{P}_{k|k-1}^{b}=\mathbf{F}_{k-1}\mathbf{P}_{k-1|k-1}^{b}\mathbf{F}%
_{k-1}^{H}+\mathbf{C}_{\mathbf{w}_{k-1}}  \label{LLMSF - pre/cor - P(k|k-1)}
\\
+\mathbf{F}_{k-1}\mathbf{C}_{\mathbf{w}_{k-1},\mathbf{x}_{k-1}}^{H}+\mathbf{C%
}_{\mathbf{w}_{k-1},\mathbf{x}_{k-1}}\mathbf{F}_{k-1}^{H}
\end{multline}%
\begin{align}
\mathbf{S}_{k|k}& =\mathbf{H}_{k}\mathbf{P}_{k|k-1}^{b}\mathbf{H}_{k}^{H}+%
\mathbf{C}_{\mathbf{v}_{k}}+\mathbf{H}_{k}\mathbf{C}_{\mathbf{v}_{k},\mathbf{%
x}_{k}}^{H}+\mathbf{C}_{\mathbf{v}_{k},\mathbf{x}_{k}}\mathbf{H}_{k}^{H}
\notag \\
\mathbf{K}_{k}^{b}& =\left( \mathbf{P}_{k|k-1}^{b}\mathbf{H}_{k}^{H}+\mathbf{%
C}_{\mathbf{v}_{k},\mathbf{x}_{k}}^{H}\right) \mathbf{S}_{k|k}^{-1}
\label{LLMSF - pre/cor - K(k)} \\
\mathbf{P}_{k|k}^{b}& =\left( \mathbf{I}-\mathbf{K}_{k}^{b}\mathbf{H}%
_{k}\right) \mathbf{P}_{k|k-1}^{b}-\mathbf{K}_{k}^{b}\mathbf{C}_{\mathbf{v}%
_{k},\mathbf{x}_{k}}  \label{LLMSF - pre/cor - P(k)}
\end{align}%
where $\mathbf{S}_{k|k}=\mathbf{C}_{\mathbf{\varepsilon }_{k}}$ and:%
\begin{equation}
\mathbf{P}_{k-1|k-1}^{b}=\underset{\left( \mathbf{G}_{k-2},\mathbf{K}%
_{k-1}\right) }{\min }\left\{ \mathbf{P}_{k-1|k-1}\left( \mathbf{G}_{k-2},%
\mathbf{K}_{k-1}\right) \right\} .  \label{recursion Pk-1|k-1}
\end{equation}%
The above recursion is also valid for $k=1$ provided that $\mathbf{P}%
_{0|0}^{b}=\mathbf{C}_{\mathbf{x}_{0}}$ and $\widehat{\mathbf{x}}_{0|0}^{b}=%
\mathbf{0}$\footnote{%
The case of a non-zero mean initial state $\mathbf{x}_{0}$ is addressed by
simply setting $\widehat{\mathbf{x}}_{0|0}=E\left[ \mathbf{x}_{0}\right] $
\cite{Kailath - Sayed - Hassibi}\cite{Simon}.} \cite{Chaumette - TAC}. As
already stressed in \cite{Chaumette - TAC}\cite{Chaumette - SPL}, the
so-called "standard LDSS model" mentioned in monographs \cite[\S 9.1]%
{Kailath - Sayed - Hassibi}\cite[\S 7.1]{Simon}\cite[\S 8.2]{Gibbs}, which
satisfies:
\end{subequations}
\begin{equation}
\begin{array}{l}
\mathbf{C}_{\mathbf{x}_{0},\mathbf{w}_{k}}=\mathbf{0},~\mathbf{C}_{\mathbf{x}%
_{0},\mathbf{v}_{k}}=\mathbf{0},~\mathbf{C}_{\mathbf{w}_{l},\mathbf{w}_{k}}=%
\mathbf{C}_{\mathbf{w}_{k}}\delta _{k}^{l}, \\
\qquad \quad ~\mathbf{C}_{\mathbf{v}_{l},\mathbf{v}_{k}}=\mathbf{C}_{\mathbf{%
v}_{k}}\delta _{k}^{l},~\mathbf{C}_{\mathbf{w}_{l},\mathbf{v}_{k}}=\mathbf{C}%
_{\mathbf{w}_{k-1},\mathbf{v}_{k}}\delta _{k}^{l+1},%
\end{array}
\label{DLS - Sequential BLE Existence Conditions - Usual}
\end{equation}%
and which has been regarded so far as leading to the general form of the KF
(without extension of the state and measurement equations) including
correlated state and measurement noise, is in fact a special case of (\ref%
{LLMSF - recursive pred/cor cond}) yielding simplified expressions of (\ref%
{LLMSF - pre/cor - P(k|k-1)}-\ref{LLMSF - pre/cor - P(k)}). However, a
thorough characterization of the subset of LDSS models compliant with (\ref%
{LLMSF - recursive pred/cor cond}) is out of the scope of the paper and is
left for future research.

\subsection{Linearly constrained LLMSE (LCLLMSE)}

The linearly constrained LLMSE is the solution of:
\begin{equation}
\mathbf{K}^{b}=\arg \underset{\mathbf{K}}{\min }\left\{ \mathbf{P}\left(
\mathbf{K}\right) \right\} \text{ s.t. }\mathbf{K\Lambda }=\mathbf{T}.
\label{LCLLMSE - def K}
\end{equation}%
To stress the fact that the LCLLMSE is different from the LLMSE, we adopt
the notation used in the deterministic framework for the MVDR estimator
(MVDRE) and its extension, aka the LCMV estimator (LCMVE) \cite[\S 6]{Van
Trees Part IV}\cite[\S 5.6]{Schreier - Scharf}. Indeed, if $\mathbf{x}$ is a
state vector and $\mathbf{y}$ is a measurement vector, one can define a
"state-former" in the same way as a beamformer in array processing or a
frequency-bin former in spectral analysis \cite[\S 6-7]{Van Trees Part IV}%
\cite[\S 5.6]{Schreier - Scharf}, that is $\mathbf{W}\in \mathcal{M}_{%
\mathbb{C}
}\left( \dim \left( \mathbf{y}\right) ,\dim \left( \mathbf{x}\right) \right)
$ yielding the state vector $\mathbf{W}^{H}\mathbf{y}$. Furthermore, this
common notation will help the reader to transpose the abundant literature on
LCMVE in the deterministic framework \cite{Vorobyov} to the stochastic
framework, since the recursive LCMVE is a special case of the recursive
LCLLMSE for LDSS models, as shown in Section \ref{S: Deterministic
parameters estimation}. All in all it simply amounts to set $\mathbf{K}=%
\mathbf{W}^{H}$. Then (\ref{LCLLMSE - def K}) becomes:
\begin{multline}
\mathbf{W}^{b}=\arg \underset{\mathbf{W}}{\min }\left\{ \mathbf{P}\left(
\mathbf{W}\right) \right\} \text{ s.t. }\mathbf{W}^{H}\mathbf{\Lambda }=%
\mathbf{T},  \label{LCLLMSE - def} \\
\mathbf{P}\left( \mathbf{W}\right) =E\left[ \left( \mathbf{W}^{H}\mathbf{y}-%
\mathbf{x}\right) \left( \mathbf{W}^{H}\mathbf{y}-\mathbf{x}\right) ^{H}%
\right] .
\end{multline}%
If $\mathbf{C}_{\mathbf{y}}$ is invertible and $\mathbf{\Lambda }$ is a full
rank matrix, it can easily be shown that \cite[(2.113)]{Diniz}:
\begin{subequations}
\begin{multline}
\mathbf{W}^{b}=\mathbf{C}_{\mathbf{y}}^{-1}\mathbf{\Lambda }\left( \mathbf{%
\Lambda }^{H}\mathbf{C}_{\mathbf{y}}^{-1}\mathbf{\Lambda }\right) ^{-1}%
\mathbf{T}^{H}+  \label{LCLLMSE - Wb} \\
\left( \mathbf{I}-\mathbf{C}_{\mathbf{y}}^{-1}\mathbf{\Lambda }\left(
\mathbf{\Lambda }^{H}\mathbf{C}_{\mathbf{y}}^{-1}\mathbf{\Lambda }\right)
^{-1}\mathbf{\Lambda }^{H}\right) \mathbb{W},
\end{multline}%
and:
\begin{multline}
\mathbf{P}\left( \mathbf{W}^{b}\right) =\mathbf{P}\left( \mathbb{W}\right) +
\label{LCLLMSE - min(P(W))} \\
\left( \mathbf{T}^{H}-\mathbf{\Lambda }^{H}\mathbb{W}\right) ^{H}\left(
\mathbf{\Lambda }^{H}\mathbf{C}_{\mathbf{y}}^{-1}\mathbf{\Lambda }\right)
^{-1}\left( \mathbf{T}^{H}-\mathbf{\Lambda }^{H}\mathbb{W}\right) ,
\end{multline}%
where $\mathbb{W}$ is the best unconstrained state-former:%
\begin{equation}
\mathbb{W}=\left( \mathbf{K}^{b}\right) ^{H}=\mathbf{C}_{\mathbf{y}}^{-1}%
\mathbf{C}_{\mathbf{y},\mathbf{x}},~\mathbf{P}\left( \mathbb{W}\right) =%
\mathbf{C}_{\mathbf{x|y}}.
\end{equation}%
The LCLLMSE coincides with the LLMSE iif: $\mathbf{T}=\mathbb{W}^{H}\mathbf{%
\Lambda }=\mathbf{K}^{b}\mathbf{\Lambda }$.

\subsection{Linear minimum variance distortionless response filter (LMVDRF)
\label{S: LMVDRF}}

Let $\overline{\mathbf{W}}_{k}=\QTATOPD[ ] {\overline{\mathbf{D}}_{k-1}}{%
\mathbf{W}_{k}}$ where $\overline{\mathbf{D}}_{k-1}\in \mathcal{M}_{%
\mathbb{C}
}\left( \mathcal{N}_{k-1},P_{k}\right) $ and $\mathbf{W}_{k}\in \mathcal{M}_{%
\mathbb{C}
}\left( N_{k},P_{k}\right) $. Since:
\end{subequations}
\begin{subequations}
\begin{equation}
\overline{\mathbf{W}}_{k}^{H}\overline{\mathbf{y}}_{k}=\left( \left(
\overline{\mathbf{W}}_{k}^{H}\overline{\mathbf{A}}_{k}\right) \mathbf{x}_{1}+%
\mathbf{G}_{k}\overline{\mathbf{w}}_{k-1}\right) +\overline{\mathbf{W}}%
_{k}^{H}\overline{\mathbf{n}}_{k}-\mathbf{G}_{k}\overline{\mathbf{w}}_{k-1},
\end{equation}%
where $\mathbf{G}_{k}\overline{\mathbf{w}}_{k-1}=\tsum_{l=1}^{k-1}\mathbf{B}%
_{k,l+1}\mathbf{w}_{l}$, $\mathbf{G}_{k}\in \mathcal{M}_{%
\mathbb{C}
}\left( P_{k},\mathcal{P}_{k-1}\right) $, $\mathcal{P}_{k}=\tsum%
\nolimits_{l=1}^{k}P_{l}$, a state-former $\overline{\mathbf{W}}_{k}$ is
distortionless iff:
\begin{equation}
\overline{\mathbf{W}}_{k}^{H}\overline{\mathbf{A}}_{k}=\mathbf{B}%
_{k,1}\Leftrightarrow \overline{\mathbf{W}}_{k}^{H}\overline{\mathbf{y}}_{k}=%
\mathbf{x}_{k}+\overline{\mathbf{W}}_{k}^{H}\overline{\mathbf{n}}_{k}-%
\mathbf{G}_{k}\overline{\mathbf{w}}_{k-1}.  \label{LDRF def}
\end{equation}%
If $\mathbf{H}_{1}$ is full rank, there exists a best distortionless
state-former in the MSE sense, aka the LMVDRF, defined by \cite{Chaumette -
SPL}:
\begin{equation}
\overline{\mathbf{W}}_{k}^{b}=\arg \underset{\overline{\mathbf{W}}_{k}}{\min
}\left\{ \mathbf{P}_{k|k}\left( \overline{\mathbf{W}}_{k}\right) \right\}
\text{ s.t. }\overline{\mathbf{W}}_{k}^{H}\overline{\mathbf{A}}_{k}=\mathbf{B%
}_{k,1},  \label{LMVDRF def}
\end{equation}%
where $\mathbf{P}_{k|k}\left( \overline{\mathbf{W}}_{k}\right) =E\left[
\left( \overline{\mathbf{W}}_{k}^{H}\overline{\mathbf{y}}_{k}-\mathbf{x}%
_{k}\right) \left( \overline{\mathbf{W}}_{k}^{H}\overline{\mathbf{y}}_{k}-%
\mathbf{x}_{k}\right) ^{H}\right] $. The MSE breakdown (\ref{MSE
factorization}-b) is also valid for any distortionless state-former,
therefore:
\end{subequations}
\begin{subequations}
\begin{gather}
\overline{\mathbf{D}}_{k-1}^{b}=\arg \underset{\overline{\mathbf{D}}_{k-1}}{%
\min }\left\{ \mathbf{Q}_{k-1}\left( \overline{\mathbf{D}}_{k-1},\mathbf{W}%
_{k}\right) \right\} \text{ s.t. }\overline{\mathbf{W}}_{k}^{H}\overline{%
\mathbf{A}}_{k}=\mathbf{B}_{k,1},  \label{D(k-1) opt} \\
\begin{array}{l}
\mathbf{Q}_{k-1}\left( \overline{\mathbf{D}}_{k-1},\mathbf{W}_{k}\right) =
\\
\qquad \quad E\left[
\begin{array}{c}
\left( \overline{\mathbf{D}}_{k-1}^{H}\overline{\mathbf{y}}_{k-1}-\left(
\mathbf{I}-\mathbf{W}_{k}^{H}\mathbf{H}_{k}\right) \mathbf{F}_{k-1}\mathbf{x}%
_{k-1}\right) \times  \\
\left( \overline{\mathbf{D}}_{k-1}^{H}\overline{\mathbf{y}}_{k-1}-\left(
\mathbf{I}-\mathbf{W}_{k}^{H}\mathbf{H}_{k}\right) \mathbf{F}_{k-1}\mathbf{x}%
_{k-1}\right) ^{H}%
\end{array}%
\right] .%
\end{array}
\notag
\end{gather}%
Furthermore, since an equivalent form of the set of linear constraints $%
\overline{\mathbf{W}}_{k}^{H}\overline{\mathbf{A}}_{k}=\mathbf{B}_{k,1}$ is:
\begin{equation}
\overline{\mathbf{D}}_{k-1}^{H}\overline{\mathbf{A}}_{k-1}=\left( \mathbf{I}-%
\mathbf{W}_{k}^{H}\mathbf{H}_{k}\right) \mathbf{F}_{k-1}\mathbf{B}_{k-1,1},
\label{distortionless constraints equivalence}
\end{equation}%
one can notice that the solution of (\ref{D(k-1) opt}) is \cite{Chaumette -
SPL}:
\begin{gather}
\overline{\mathbf{D}}_{k-1}^{b}=\overline{\mathbf{W}}_{k-1}^{b}\left( \left(
\mathbf{I}-\mathbf{W}_{k}^{H}\mathbf{H}_{k}\right) \mathbf{F}_{k-1}\right)
^{H},\qquad \qquad   \label{D(k-1) opt - equiv} \\
\overline{\mathbf{W}}_{k-1}^{b}=\arg \underset{\overline{\mathbf{W}}_{k-1}}{%
\min }\left\{ \mathbf{P}_{k-1|k-1}\left( \overline{\mathbf{W}}_{k-1}\right)
\right\} \qquad \qquad \qquad ~  \notag \\
\qquad \qquad \qquad \qquad \qquad \text{ s.t. }\overline{\mathbf{W}}%
_{k-1}^{H}\overline{\mathbf{A}}_{k-1}=\mathbf{B}_{k-1,1},
\end{gather}%
provided that $\mathbf{C}_{\overline{\mathbf{n}}_{k-1}}$ is invertible, and
yields:%
\begin{multline}
\mathbf{Q}_{k-1}\left( \overline{\mathbf{D}}_{k-1}^{b},\mathbf{W}_{k}\right)
=\left( \mathbf{I}-\mathbf{W}_{k}^{H}\mathbf{H}_{k}\right) \mathbf{F}%
_{k-1}\times  \\
\mathbf{P}_{k-1|k-1}^{b}\mathbf{F}_{k-1}^{H}\left( \mathbf{I}-\mathbf{W}%
_{k}^{H}\mathbf{H}_{k}\right) ^{H},
\end{multline}%
where $\mathbf{P}_{k-1|k-1}^{b}=\mathbf{P}_{k-1|k-1}\left( \overline{\mathbf{%
W}}_{k-1}^{b}\right) $. Finally, $\forall k\geq 2$, the MSE breakdown (\ref%
{MSE factorization}-b) $\mathbf{P}_{k|k}\left( \overline{\mathbf{D}}%
_{k-1}^{b},\mathbf{W}_{k}\right) $ shares the same general form of the
Joseph stabilized version of the covariance measurement update equation (\ref%
{Joseph}), provided that one substitutes $\mathbf{W}_{k}^{H}$ for $\mathbf{K}%
_{k}$. Therefore, if $\mathbf{H}_{1}$ is full rank, the LMVDRF shares the
same recursion as the KF:
\end{subequations}
\begin{equation}
\widehat{\mathbf{x}}_{k|k}^{b}=\mathbf{F}_{k-1}\widehat{\mathbf{x}}%
_{k-1|k-1}^{b}+\left( \mathbf{W}_{k}^{b}\right) ^{H}\left( \mathbf{y}_{k}-%
\mathbf{H}_{k}\mathbf{F}_{k-1}\widehat{\mathbf{x}}_{k-1|k-1}^{b}\right) ,
\label{recursion LMVDRF}
\end{equation}%
where $\mathbf{W}_{k}^{b}$ is updated according to (\ref{LLMSF - pre/cor -
P(k|k-1)}-c) provided that one substitutes $\left( \mathbf{W}_{k}^{b}\right)
^{H}$ for $\mathbf{K}_{k}^{b}$, except at time $k=1$ where, if $\mathbf{C}_{%
\mathbf{v}_{1}}$ is full rank:%
\begin{multline*}
\widehat{\mathbf{x}}_{1|1}^{b}=\left( \mathbf{W}_{1}^{b}\right) ^{H}\mathbf{y%
}_{1},~\mathbf{W}_{1}^{b}=\mathbf{C}_{\mathbf{v}_{1}}^{-1}\mathbf{H}%
_{1}\left( \mathbf{H}_{1}^{H}\mathbf{C}_{\mathbf{v}_{1}}^{-1}\mathbf{H}%
_{1}\right) ^{-1}, \\
\mathbf{P}_{1|1}^{b}=\left( \mathbf{H}_{1}^{H}\mathbf{C}_{\mathbf{v}%
_{1}}^{-1}\mathbf{H}_{1}\right) ^{-1}.
\end{multline*}

\section{Linearly constrained KF for LDSS models}

A linearly constrained\ LLMSF (LCLLMSF) is the solution of:
\begin{equation}
\overline{\mathbf{W}}_{k}^{b}=\arg \underset{\overline{\mathbf{W}}_{k}}{\min
}\left\{ \mathbf{P}_{k|k}\left( \overline{\mathbf{W}}_{k}\right) \right\}
\text{ s.t. }\overline{\mathbf{W}}_{k}^{H}\mathbf{\Lambda }_{k}=\mathbf{%
\Gamma }_{k}.  \label{LCLLMSF def}
\end{equation}%
In order to make use of (\ref{LCLLMSE - Wb}-\ref{LCLLMSE - min(P(W))}), we
limit ourselves to the case where $\mathbf{\Lambda }_{k}$ is full rank and $%
\mathbf{C}_{\overline{\mathbf{y}}_{l}}$, $1\leq l\leq k$, are invertible. As
shown in Subsection \ref{S: LMVDRF}, the LMVDRF is an example of LCLLMSF
(obtained where $\mathbf{\Lambda }_{k}=\overline{\mathbf{A}}_{k}$ and $%
\mathbf{\Gamma }_{k}=\mathbf{B}_{k,1}$) with a recursive predictor/corrector
format (\ref{recursion LMVDRF}). From the derivation of the LMVDRF outlined
above, a sensible generalization of the set of constraints (\ref%
{distortionless constraints equivalence}) compatible with the
predictor/corrector recursion is the set $\mathcal{C}_{k}^{1}$, $k\geq 2$,
defined by:
\begin{subequations}
\begin{equation}
\mathcal{C}_{k}^{1}:\overline{\mathbf{W}}_{k}^{H}\left[
\begin{array}{cc}
\mathbf{\Lambda }_{k-1} & \mathbf{0} \\
\mathbf{H}_{k}\mathbf{F}_{k-1}\mathbf{\Gamma }_{k-1} & \mathbf{\Delta }_{k}%
\end{array}%
\right] =\left[ \mathbf{F}_{k-1}\mathbf{\Gamma }_{k-1}~\mathbf{T}_{k}\right]
,  \label{Constraints C1}
\end{equation}%
that is:%
\begin{align}
\mathcal{C}_{k}^{1,1}& :\overline{\mathbf{D}}_{k-1}^{H}\mathbf{\Lambda }%
_{k-1}=\left( \mathbf{I}-\mathbf{W}_{k}^{H}\mathbf{H}_{k}\right) \mathbf{F}%
_{k-1}\mathbf{\Gamma }_{k-1},  \label{constraints C1 - Dk-1} \\
\mathcal{C}_{k}^{1,2}& :\quad ~\mathbf{W}_{k}^{H}\mathbf{\Delta }_{k}=%
\mathbf{T}_{k},  \label{constraints  C1 -  Wk}
\end{align}%
where both $\mathbf{\Lambda }_{k-1}$ and $\mathbf{\Delta }_{k}$ are full
rank. Indeed, since the MSE breakdown (\ref{MSE factorization}-b) is valid
for any state-former, therefore under (\ref{Constraints C1}):
\end{subequations}
\begin{multline}
\overline{\mathbf{D}}_{k-1}^{b}=\arg \underset{\overline{\mathbf{D}}_{k-1}}{%
\min }\left\{ \mathbf{Q}_{k-1}\left( \overline{\mathbf{D}}_{k-1},\mathbf{W}%
_{k}\right) \right\}   \notag \\
\text{ s.t. }\overline{\mathbf{D}}_{k-1}^{H}\mathbf{\Lambda }_{k-1}=\left(
\mathbf{I}-\mathbf{W}_{k}^{H}\mathbf{H}_{k}\right) \mathbf{F}_{k-1}\mathbf{%
\Gamma }_{k-1},
\end{multline}%
that is according to (\ref{LCLLMSE - Wb}):
\begin{subequations}
\begin{gather}
\quad \overline{\mathbf{D}}_{k-1}^{b}=\overline{\mathbf{W}}_{k-1}^{b}\left(
\left( \mathbf{I}-\mathbf{W}_{k}^{H}\mathbf{H}_{k}\right) \mathbf{F}%
_{k-1}\right) ^{H},\qquad   \label{Dk-1 / Wk-1} \\
\overline{\mathbf{W}}_{k-1}^{b}=\arg \underset{\overline{\mathbf{W}}_{k-1}}{%
\min }\left\{ \mathbf{P}_{k-1|k-1}\left( \overline{\mathbf{W}}_{k-1}\right)
\right\} \qquad \qquad   \notag \\
\qquad \qquad \qquad \qquad \text{ s.t. }\overline{\mathbf{W}}_{k-1}^{H}%
\mathbf{\Lambda }_{k-1}=\mathbf{\Gamma }_{k-1},
\end{gather}%
and yielding:%
\begin{multline}
\mathbf{Q}_{k-1}\left( \overline{\mathbf{D}}_{k-1}^{b},\mathbf{W}_{k}\right)
=\left( \mathbf{I}-\mathbf{W}_{k}^{H}\mathbf{H}_{k}\right) \times  \\
\mathbf{F}_{k-1}\mathbf{P}_{k-1|k-1}^{b}\mathbf{F}_{k-1}^{H}\left( \mathbf{I}%
-\mathbf{W}_{k}^{H}\mathbf{H}_{k}\right) ^{H}.
\end{multline}%
It is then worth noticing that, likewise:
\end{subequations}
\begin{multline*}
\overline{\mathbf{W}}_{k-1}^{b}\mathbf{F}_{k-1}^{H}= \\
\arg \underset{\overline{\mathbf{W}}_{k-1}}{\min }\left\{ E\left[
\begin{array}{c}
\left( \overline{\mathbf{W}}_{k-1}^{H}\overline{\mathbf{y}}_{k-1}-\mathbf{F}%
_{k-1}\mathbf{x}_{k-1}\right) \times  \\
\left( \overline{\mathbf{W}}_{k-1}^{H}\overline{\mathbf{y}}_{k-1}-\mathbf{F}%
_{k-1}\mathbf{x}_{k-1}\right) ^{H}%
\end{array}%
\right] \right\} \text{ } \\
\text{s.t. }\overline{\mathbf{W}}_{k-1}^{H}\mathbf{\Lambda }_{k-1}=\mathbf{F}%
_{k-1}\mathbf{\Gamma }_{k-1},
\end{multline*}%
which, under (\ref{LLMSF - recursive pred/cor cond}), can be rewritten as:%
\begin{multline}
\overline{\mathbf{W}}_{k-1}^{b}\mathbf{F}_{k-1}^{H}=  \label{xk|k-1 - def} \\
\arg \underset{\overline{\mathbf{W}}_{k-1}}{\min }\left\{ E\left[
\begin{array}{c}
\left( \overline{\mathbf{W}}_{k-1}^{H}\overline{\mathbf{y}}_{k-1}-\mathbf{x}%
_{k}\right) \times  \\
\left( \overline{\mathbf{W}}_{k-1}^{H}\overline{\mathbf{y}}_{k-1}-\mathbf{x}%
_{k}\right) ^{H}%
\end{array}%
\right] \right\} \text{ } \\
\text{s.t. }\overline{\mathbf{W}}_{k-1}^{H}\mathbf{\Lambda }_{k-1}=\mathbf{F}%
_{k-1}\mathbf{\Gamma }_{k-1},
\end{multline}%
since then:%
\begin{multline*}
E\left[
\begin{array}{c}
\left( \overline{\mathbf{W}}_{k-1}^{H}\overline{\mathbf{y}}_{k-1}-\mathbf{x}%
_{k}\right) \times  \\
\left( \overline{\mathbf{W}}_{k-1}^{H}\overline{\mathbf{y}}_{k-1}-\mathbf{x}%
_{k}\right) ^{H}%
\end{array}%
\right] = \\
E\left[
\begin{array}{c}
\left( \overline{\mathbf{W}}_{k-1}^{H}\overline{\mathbf{y}}_{k-1}-\mathbf{F}%
_{k-1}\mathbf{x}_{k-1}\right) \times  \\
\left( \overline{\mathbf{W}}_{k-1}^{H}\overline{\mathbf{y}}_{k-1}-\mathbf{F}%
_{k-1}\mathbf{x}_{k-1}\right) ^{H}%
\end{array}%
\right]  \\
+\mathbf{C}_{\mathbf{w}_{k-1}}+\mathbf{F}_{k-1}\mathbf{C}_{\mathbf{x}_{k-1},%
\mathbf{w}_{k-1}}+\mathbf{C}_{\mathbf{w}_{k-1},\mathbf{x}_{k-1}}\mathbf{F}%
_{k-1}^{H}.
\end{multline*}%
Thus:
\begin{subequations}
\begin{align}
\widehat{\mathbf{x}}_{k|k-1}^{b}& =\mathbf{F}_{k-1}\widehat{\mathbf{x}}%
_{k-1|k-1}^{b}, \\
\mathbf{P}_{k|k-1}^{b}& =\mathbf{F}_{k-1}\mathbf{P}_{k-1|k-1}^{b}\mathbf{F}%
_{k-1}^{H}+\mathbf{C}_{\mathbf{w}_{k-1}}  \notag \\
& +\mathbf{F}_{k-1}\mathbf{C}_{\mathbf{x}_{k-1},\mathbf{w}_{k-1}}+\mathbf{C}%
_{\mathbf{w}_{k-1},\mathbf{x}_{k-1}}\mathbf{F}_{k-1}^{H},
\end{align}%
where $\widehat{\mathbf{x}}_{k|k-1}^{b}$ is the solution of (\ref{xk|k-1 -
def}). Therefore, under (\ref{Constraints C1}), the MSE breakdown (\ref{MSE
factorization}-b) $\mathbf{P}_{k|k}\left( \overline{\mathbf{D}}_{k-1}^{b},%
\mathbf{W}_{k}\right) $ yields the general form of the linearly constrained
Joseph stabilized version of the covariance measurement update equation (\ref%
{Joseph}):
\end{subequations}
\begin{multline}
\mathbf{W}_{k}^{b}=\arg \underset{\mathbf{W}_{k}}{\min }\left\{ E\left[
\begin{array}{c}
\left( \mathbf{W}_{k}^{H}\mathbf{\varepsilon }_{k}-\left( \mathbf{x}_{k}-%
\widehat{\mathbf{x}}_{k|k-1}^{b}\right) \right) \times  \\
\left( \mathbf{W}_{k}^{H}\mathbf{\varepsilon }_{k}-\left( \mathbf{x}_{k}-%
\widehat{\mathbf{x}}_{k|k-1}^{b}\right) \right) ^{H}%
\end{array}%
\right] \right\} \text{ }  \label{Linearly Constrained Joseph} \\
\text{s.t. }\mathbf{W}_{k}^{H}\mathbf{\Delta }_{k}=\mathbf{T}_{k},
\end{multline}%
where $\mathbf{\varepsilon }_{k}$ still stands for the innovations vector (%
\ref{innovations vector}) since (from a similar derivation as the one for $%
\widehat{\mathbf{x}}_{k|k-1}^{b}$):%
\begin{multline}
\overline{\mathbf{W}}_{k-1}^{b}\mathbf{F}_{k-1}^{H}\mathbf{H}_{k}^{H}= \\
\arg \underset{\overline{\mathbf{W}}_{k-1}}{\min }\left\{ E\left[
\begin{array}{c}
\left( \overline{\mathbf{W}}_{k-1}^{H}\overline{\mathbf{y}}_{k-1}-\mathbf{y}%
_{k}\right) \times  \\
\left( \overline{\mathbf{W}}_{k-1}^{H}\overline{\mathbf{y}}_{k-1}-\mathbf{y}%
_{k}\right) ^{H}%
\end{array}%
\right] \right\} \text{ } \\
\text{s.t. }\overline{\mathbf{W}}_{k-1}^{H}\mathbf{\Lambda }_{k-1}=\mathbf{H}%
_{k}\mathbf{F}_{k-1}\mathbf{\Gamma }_{k-1}.
\end{multline}%
The solution of (\ref{Linearly Constrained Joseph}) is given by (\ref%
{LCLLMSE - Wb}-\ref{LCLLMSE - min(P(W))}), which yields the following
linearly constrained KF (LCKF) recursion at time $k$:
\begin{subequations}
\begin{equation}
\widehat{\mathbf{x}}_{k|k}^{b}=\mathbf{F}_{k-1}\widehat{\mathbf{x}}%
_{k-1|k-1}^{b}+\left( \mathbf{W}_{k}^{b}\right) ^{H}\left( \mathbf{y}_{k}-%
\mathbf{H}_{k}\mathbf{F}_{k-1}\widehat{\mathbf{x}}_{k-1|k-1}^{b}\right) ,
\label{recursion C1 - beg}
\end{equation}%
\begin{multline}
\mathbf{P}_{k|k-1}^{b}=\mathbf{F}_{k-1}\mathbf{P}_{k-1|k-1}^{b}\mathbf{F}%
_{k-1}^{H}+\mathbf{C}_{\mathbf{w}_{k-1}}+ \\
\mathbf{F}_{k-1}\mathbf{C}_{\mathbf{x}_{k-1},\mathbf{w}_{k-1}}+\mathbf{C}_{%
\mathbf{w}_{k-1},\mathbf{x}_{k-1}}\mathbf{F}_{k-1}^{H}
\end{multline}%
\begin{align}
\mathbf{S}_{k|k}& =\mathbf{H}_{k}\mathbf{P}_{k|k-1}^{b}\mathbf{H}_{k}^{H}+%
\mathbf{C}_{\mathbf{v}_{k}}+\mathbf{H}_{k}\mathbf{C}_{\mathbf{x}_{k},\mathbf{%
v}_{k}}+\mathbf{C}_{\mathbf{v}_{k},\mathbf{x}_{k}}\mathbf{H}_{k}^{H} \\
\mathbb{W}_{k}& =\mathbf{S}_{k|k}^{-1}\left( \mathbf{H}_{k}\mathbf{P}%
_{k|k-1}^{b}+\mathbf{C}_{\mathbf{v}_{k},\mathbf{x}_{k}}\right)  \\
\mathbf{W}_{k}^{b}& =\mathbb{W}_{k}+\mathbf{S}_{k|k}^{-1}\mathbf{\Delta }%
_{k}\left( \mathbf{\Delta }_{k}^{H}\mathbf{S}_{k|k}^{-1}\mathbf{\Delta }%
_{k}\right) ^{-1}\left( \mathbf{T}_{k}-\mathbb{W}_{k}^{H}\mathbf{\Delta }%
_{k}\right) ^{H}
\end{align}%
\begin{multline}
\mathbf{P}_{k|k}^{b}=\left( \mathbf{I}-\mathbb{W}_{k}^{H}\mathbf{H}%
_{k}\right) \mathbf{P}_{k|k-1}^{b}-\mathbb{W}_{k}^{H}\mathbf{C}_{\mathbf{v}%
_{k},\mathbf{x}_{k}}+  \label{recursion C1 - end} \\
\left( \mathbf{T}_{k}-\mathbb{W}_{k}^{H}\mathbf{\Delta }_{k}\right) \left(
\mathbf{\Delta }_{k}^{H}\mathbf{S}_{k|k}^{-1}\mathbf{\Delta }_{k}\right)
^{-1}\left( \mathbf{T}_{k}-\mathbb{W}_{k}^{H}\mathbf{\Delta }_{k}\right) ^{H}
\end{multline}%
and:
\begin{gather}
\mathbf{P}_{k-1|k-1}^{b}=\underset{\overline{\mathbf{W}}_{k-1}}{\min }%
\left\{ \mathbf{P}_{k-1|k-1}\left( \overline{\mathbf{W}}_{k-1}\right)
\right\} \qquad \qquad \qquad ~  \notag \\
\qquad \qquad \qquad \qquad \qquad \text{ s.t. }\overline{\mathbf{W}}%
_{k-1}^{H}\mathbf{\Lambda }_{k-1}=\mathbf{\Gamma }_{k-1}.
\label{recursion C1 - Pk-1|k-1}
\end{gather}

\subsection{The general case}

In the general case, we look for the solution of (\ref{LCLLMSF def}) where:
\end{subequations}
\begin{equation}
\left[
\begin{array}{c}
\overline{\mathbf{D}}_{k-1} \\
\mathbf{W}_{k}%
\end{array}%
\right] ^{H}\left[
\begin{array}{c}
\mathbf{\Phi }_{k-1} \\
\mathbf{\Psi }_{k}%
\end{array}%
\right] =\mathbf{\Gamma }_{k}\Leftrightarrow \overline{\mathbf{D}}_{k-1}^{H}%
\mathbf{\Phi }_{k-1}=\mathbf{\Gamma }_{k}-\mathbf{W}_{k}^{H}\mathbf{\Psi }%
_{k}.  \label{Constraints general}
\end{equation}%
Since the MSE breakdown (\ref{MSE factorization}-b) is valid for any
state-former, therefore under (\ref{Constraints general}):%
\begin{multline}
\overline{\mathbf{D}}_{k-1}^{b}=\arg \underset{\overline{\mathbf{D}}_{k-1}}{%
\min }\left\{ \mathbf{Q}_{k-1}\left( \overline{\mathbf{D}}_{k-1},\mathbf{W}%
_{k}\right) \right\}  \notag \\
\text{ s.t. }\overline{\mathbf{D}}_{k-1}^{H}\mathbf{\Phi }_{k-1}=\mathbf{%
\Gamma }_{k}-\mathbf{W}_{k}^{H}\mathbf{\Psi }_{k}.
\end{multline}%
Provided that $\mathbf{\Phi }_{k-1}$ is full rank, then according to (\ref%
{LCLLMSE - Wb}-\ref{LCLLMSE - min(P(W))}):%
\begin{multline}
\overline{\mathbf{D}}_{k-1}^{b}=  \label{Db(k-1) general} \\
\mathbf{C}_{\overline{\mathbf{y}}_{k-1}}^{-1}\mathbf{\Phi }_{k-1}\left(
\mathbf{\Phi }_{k-1}^{H}\mathbf{C}_{\overline{\mathbf{y}}_{k-1}}^{-1}\mathbf{%
\Phi }_{k-1}\right) ^{-1}\left( \mathbf{\Gamma }_{k}-\mathbf{W}_{k}^{H}%
\mathbf{\Psi }_{k}\right) ^{H} \\
+\left( \mathbf{I}-\mathbf{C}_{\overline{\mathbf{y}}_{k-1}}^{-1}\mathbf{\Phi
}_{k-1}\left( \mathbf{\Phi }_{k-1}^{H}\mathbf{C}_{\overline{\mathbf{y}}%
_{k-1}}^{-1}\mathbf{\Phi }_{k-1}\right) ^{-1}\mathbf{\Phi }_{k-1}^{H}\right)
\overline{\mathbb{W}}_{k-1}^{b} \\
\times \left( \left( \mathbf{I}-\mathbf{W}_{k}^{H}\mathbf{H}_{k}\right)
\mathbf{F}_{k-1}\right) ^{H},
\end{multline}%
where $\overline{\mathbb{W}}_{k-1}^{b}=\mathbf{C}_{\overline{\mathbf{y}}%
_{k-1}}^{-1}\mathbf{C}_{\overline{\mathbf{y}}_{k-1},\mathbf{x}_{k-1}}$ is
the best unconstrained state-former. It is noteworthy that (\ref{Db(k-1)
general}) can be recasted as:
\begin{subequations}
\begin{multline*}
\overline{\mathbf{D}}_{k-1}^{b}=\overline{\mathbf{W}}_{k-1}^{b}\left( \left(
\mathbf{I}-\mathbf{W}_{k}^{H}\mathbf{H}_{k}\right) \mathbf{F}_{k-1}\right)
^{H} \\
+\mathbf{C}_{\overline{\mathbf{y}}_{k-1}}^{-1}\mathbf{\Phi }_{k-1}\left(
\mathbf{\Phi }_{k-1}^{H}\mathbf{C}_{\overline{\mathbf{y}}_{k-1}}^{-1}\mathbf{%
\Phi }_{k-1}\right) ^{-1}\times \\
\left( \mathbf{\Gamma }_{k}-\mathbf{W}_{k}^{H}\mathbf{\Psi }_{k}-\left(
\mathbf{I}-\mathbf{W}_{k}^{H}\mathbf{H}_{k}\right) \mathbf{F}_{k-1}\mathbf{%
\Gamma }_{k-1}\right) ^{H},
\end{multline*}%
where:
\end{subequations}
\begin{gather*}
\overline{\mathbf{W}}_{k-1}^{b}=\arg \underset{\overline{\mathbf{W}}_{k-1}}{%
\min }\left\{ \mathbf{P}_{k-1|k-1}\left( \overline{\mathbf{W}}_{k-1}\right)
\right\} \qquad \qquad \qquad ~ \\
\qquad \qquad \qquad \qquad \qquad \text{ s.t. }\overline{\mathbf{W}}%
_{k-1}^{H}\mathbf{\Phi }_{k-1}=\mathbf{\Gamma }_{k-1}.
\end{gather*}%
Therefore, the solution of (\ref{LCLLMSF def}) follows a predictor/corrector
recursion (\ref{recursion C1 - beg}) iff:%
\begin{equation*}
\mathbf{\Gamma }_{k}-\mathbf{W}_{k}^{H}\mathbf{\Psi }_{k}-\left( \mathbf{I}-%
\mathbf{W}_{k}^{H}\mathbf{H}_{k}\right) \mathbf{F}_{k-1}\mathbf{\Gamma }%
_{k-1}=\mathbf{0},
\end{equation*}%
that is iff:%
\begin{equation*}
\overline{\mathbf{D}}_{k-1}^{H}\mathbf{\Phi }_{k-1}=\left( \mathbf{I}-%
\mathbf{W}_{k}^{H}\mathbf{H}_{k}\right) \mathbf{F}_{k-1}\mathbf{\Gamma }%
_{k-1},
\end{equation*}%
which is $\mathcal{C}_{k}^{1,1}$ (\ref{constraints C1 - Dk-1}). As a
consequence, the most general form of (\ref{Constraints general}) leading to
a solution following a predictor/corrector recursion is $\mathcal{C}_{k}^{1}$
(\ref{Constraints C1}).

\subsection{Constraints variants}

Obviously the set of constraints $\mathcal{C}_{k}^{2}$ defined as the
restriction of $\mathcal{C}_{k}^{1}$ (\ref{Constraints C1}) to $\mathcal{C}%
_{k}^{1,2}$ (\ref{constraints C1 - Wk}):
\begin{subequations}
\begin{equation}
\mathcal{C}_{k}^{2}:\overline{\mathbf{W}}_{k}^{H}\left[
\begin{array}{c}
\mathbf{0} \\
\mathbf{\Delta }_{k}%
\end{array}%
\right] =\mathbf{T}_{k}~\Leftrightarrow ~\mathbf{W}_{k}^{H}\mathbf{\Delta }%
_{k}=\mathbf{T}_{k},  \label{Constraints C2}
\end{equation}%
follows as well the recursion (\ref{recursion C1 - beg}-\ref{recursion C1 -
end}), except that (\ref{recursion C1 - Pk-1|k-1}) must be replaced with:%
\begin{equation}
\mathbf{P}_{k-1|k-1}^{b}=\underset{\overline{\mathbf{W}}_{k-1}}{\min }%
\left\{ \mathbf{P}_{k-1|k-1}\left( \overline{\mathbf{W}}_{k-1}\right)
\right\} ,  \label{recursion C2 - Pk-1|k-1}
\end{equation}%
which means that $\overline{\mathbf{W}}_{k-1}$ is unconstrained. In the same
vein, the set of constraints $\mathcal{C}_{k}^{3}$ defined as the
restriction of $\mathcal{C}_{k}^{1}$ (\ref{Constraints C1}) to $\mathcal{C}%
_{k}^{1,1}$ (\ref{constraints C1 - Dk-1}):
\end{subequations}
\begin{subequations}
\begin{multline}
\mathcal{C}_{k}^{3}:\overline{\mathbf{W}}_{k}^{H}\left[
\begin{array}{c}
\mathbf{\Lambda }_{k-1} \\
\mathbf{H}_{k}\mathbf{F}_{k-1}\mathbf{\Gamma }_{k-1}%
\end{array}%
\right] =\mathbf{F}_{k-1}\mathbf{\Gamma }_{k-1}  \label{Constraints C3} \\
\Leftrightarrow \overline{\mathbf{D}}_{k-1}^{H}\mathbf{\Lambda }%
_{k-1}=\left( \mathbf{I}-\mathbf{W}_{k}^{H}\mathbf{H}_{k}\right) \mathbf{F}%
_{k-1}\mathbf{\Gamma }_{k-1},
\end{multline}%
follows the standard recursion (\ref{LLMSF - pre/cor - P(k|k-1)}-\ref{LLMSF
- pre/cor - P(k)}) provided that one substitutes $\left( \mathbf{W}%
_{k}^{b}\right) ^{H}$ for $\mathbf{K}_{k}^{b}$ and that $\mathbf{P}%
_{k-1|k-1}^{b}$ is the solution of (\ref{recursion C1 - Pk-1|k-1}):%
\begin{multline}
\mathbf{P}_{k-1|k-1}^{b}=\underset{\overline{\mathbf{W}}_{k-1}}{\min }%
\left\{ \mathbf{P}_{k-1|k-1}\left( \overline{\mathbf{W}}_{k-1}\right)
\right\}  \label{recursion C3 - Pk-1|k-1} \\
\text{s.t. }\overline{\mathbf{W}}_{k-1}^{H}\mathbf{\Lambda }_{k-1}=\mathbf{%
\Gamma }_{k-1}.
\end{multline}

\subsection{Constraints combination}

Actually, the introduction of the first set of constraints at a given time $%
k $ is provided by:\newline
$\bullet $ $\mathcal{C}_{k}^{3}$ if $k=2$, since then $\mathbf{P}%
_{k-1|k-1}^{b}$ (\ref{recursion C3 - Pk-1|k-1}) results from a LCKF at time $%
k=1$.\newline
$\bullet $ $\mathcal{C}_{k}^{2}$ if $k>2$, since then $\mathbf{P}%
_{k-1|k-1}^{b}$ (\ref{recursion C2 - Pk-1|k-1}) results from an
unconstrained KF at time $k-1$.\newline
For time $k+1$, $\mathcal{C}_{k}^{2}/\mathcal{C}_{2}^{3}$ propagates via (%
\ref{Dk-1 / Wk-1}) either in the form of $\mathcal{C}_{k+1}^{3}$ leading to:
\end{subequations}
\begin{equation*}
\left[
\begin{array}{c}
\overline{\mathbf{D}}_{k} \\
\mathbf{W}_{k+1}%
\end{array}%
\right] ^{H}\left[
\begin{array}{c}
\left[
\begin{array}{c}
\mathbf{0} \\
\mathbf{\Delta }_{k}%
\end{array}%
\right] \\
\mathbf{H}_{k+1}\mathbf{F}_{k}\mathbf{T}_{k}%
\end{array}%
\right] =\mathbf{F}_{k}\mathbf{T}_{k},
\end{equation*}%
or in the form of $\mathcal{C}_{k+1}^{1}$ leading to:%
\begin{equation*}
\left[
\begin{array}{c}
\overline{\mathbf{D}}_{k} \\
\mathbf{W}_{k+1}%
\end{array}%
\right] ^{H}\left[
\begin{array}{cc}
\left[
\begin{array}{c}
\mathbf{0} \\
\mathbf{\Delta }_{k}%
\end{array}%
\right] & \mathbf{0} \\
\mathbf{H}_{k+1}\mathbf{F}_{k}\mathbf{T}_{k} & \mathbf{\Delta }_{k+1}%
\end{array}%
\right] =\left[ \mathbf{F}_{k}\mathbf{T}_{k}~\mathbf{T}_{k+1}\right] .
\end{equation*}%
For instance, the set of linear constraints (\ref{LCLLMSF def}) associated
to the sequence $\left\{ \mathcal{C}_{2}^{3},\ldots ,\mathcal{C}%
_{k}^{3}\right\} $ is characterized by:
\begin{subequations}
\begin{equation}
\mathbf{\Lambda }_{k}=\left[
\begin{array}{c}
\mathbf{\Delta }_{1} \\
\mathbf{H}_{2}\mathbf{B}_{2,1}\mathbf{T}_{1} \\
\vdots \\
\mathbf{H}_{k-1}\mathbf{B}_{k-1,1}\mathbf{T}_{1} \\
\mathbf{H}_{k}\mathbf{B}_{k,1}\mathbf{T}_{1}%
\end{array}%
\right] ,~\mathbf{\Gamma }_{k}=\mathbf{B}_{k,1}\mathbf{T}_{1}.
\label{Constraints combination - Ex1}
\end{equation}%
If $\mathbf{\Delta }_{1}=\mathbf{H}_{1}$ and $\mathbf{T}_{1}=\mathbf{I}$,
then $\mathbf{\Lambda }_{k}=\overline{\mathbf{A}}_{k}$ and $\mathbf{\Gamma }%
_{k}=\mathbf{B}_{k,1}$, which means that the sequence $\left\{ \mathcal{C}%
_{2}^{3},\ldots ,\mathcal{C}_{k}^{3}\right\} $ yields the LMVDRF. Another
example is given by the set of constraints (\ref{LCLLMSF def}) associated to
the sequence $\left\{ \mathcal{C}_{2}^{1},\ldots ,\mathcal{C}%
_{k}^{1}\right\} $ which is characterized by:
\begin{multline}
\mathbf{\Lambda }_{k}=  \label{Constraints combination - Ex2} \\
\left[
\begin{array}{ccccc}
\mathbf{\Delta }_{1} & \mathbf{0} & \mathbf{0} & \mathbf{0} & \mathbf{0} \\
\mathbf{H}_{2}\mathbf{B}_{2,1}\mathbf{T}_{1} & \mathbf{\Delta }_{2} &
\mathbf{0} & \mathbf{0} & \mathbf{0} \\
\vdots & \vdots & \ddots & \mathbf{0} & \mathbf{0} \\
\mathbf{H}_{k-1}\mathbf{B}_{k-1,1}\mathbf{T}_{1} & \mathbf{H}_{k-1}\mathbf{B}%
_{k-1,2}\mathbf{T}_{2} & \ldots & \mathbf{\Delta }_{k-1} & \mathbf{0} \\
\mathbf{H}_{k}\mathbf{B}_{k,1}\mathbf{T}_{1} & \mathbf{H}_{k}\mathbf{B}_{k,2}%
\mathbf{T}_{2} & \ldots & \mathbf{H}_{k}\mathbf{F}_{k-1}\mathbf{T}_{k-1} &
\mathbf{\Delta }_{k}%
\end{array}%
\right] ,\quad \\
\mathbf{\Gamma }_{k}=\left[
\begin{array}{ccccc}
\mathbf{B}_{k,1}\mathbf{T}_{1} & \mathbf{B}_{k,2}\mathbf{T}_{2} & \ldots &
\mathbf{B}_{k,k-1}\mathbf{T}_{k-1} & \mathbf{T}_{k}%
\end{array}%
\right] .
\end{multline}%
Looking at (\ref{Constraints combination - Ex1}) and (\ref{Constraints
combination - Ex2}), it seems difficult to have a clear understanding of the
equivalent system of constraints, i.e. $\overline{\mathbf{W}}_{k}^{H}\mathbf{%
\Lambda }_{k}=\mathbf{\Gamma }_{k}$, associated with any combination of $%
\left\{ \mathcal{C}_{l}^{i_{l}},\ldots ,\mathcal{C}_{l^{\prime
}}^{i_{l^{\prime }}}\right\} $, $i_{l},i_{l^{\prime }}\in \left\{
1,2,3\right\} $, $2\leq l\leq l^{\prime }\leq k$. However two properties of
the predictor/corrector recursion (\ref{recursion C1 - beg}) are worth
knowing in order to grasp the general effect of some constraints. First, it
is known that rewriting (\ref{recursion C1 - beg}) as:
\end{subequations}
\begin{multline}
\widehat{\mathbf{x}}_{k|k}^{b}-\mathbf{x}_{k}=\left( \mathbf{I}-\left(
\mathbf{W}_{k}^{b}\right) ^{H}\mathbf{H}_{k}\right) \mathbf{F}_{k-1}\left(
\widehat{\mathbf{x}}_{k-1|k-1}^{b}-\mathbf{x}_{k-1}\right)
\label{x(k|k) - xk} \\
-\left( \mathbf{I}-\left( \mathbf{W}_{k}^{b}\right) ^{H}\mathbf{H}%
_{k}\right) \mathbf{w}_{k-1}+\left( \mathbf{W}_{k}^{b}\right) ^{H}\mathbf{v}%
_{k},
\end{multline}%
allows to prove the unbiasedness property propagation:%
\begin{equation}
\forall \mathbf{W}_{k}^{b}:E\left[ \widehat{\mathbf{x}}_{k-1|k-1}^{b}-%
\mathbf{x}_{k-1}\right] =\mathbf{0~\Rightarrow ~}E\left[ \widehat{\mathbf{x}}%
_{k|k}^{b}-\mathbf{x}_{k}\right] =\mathbf{0}.
\label{unbiasedness propagation}
\end{equation}%
Second, if $\widehat{\mathbf{x}}_{k-1|k-1}^{b}$ is a linear distortionless
response filter, i.e. (\ref{LDRF def}):
\begin{multline*}
\widehat{\mathbf{x}}_{k-1|k-1}^{b}=\left( \overline{\mathbf{W}}%
_{k-1}^{b}\right) ^{H}\overline{\mathbf{y}}_{k-1} \\
=\mathbf{x}_{k-1}+\left( \overline{\mathbf{W}}_{k-1}^{b}\right) ^{H}%
\overline{\mathbf{n}}_{k-1}-\mathbf{G}_{k-1}\overline{\mathbf{w}}_{k-2,}
\end{multline*}%
then (\ref{x(k|k) - xk}) becomes:%
\begin{multline*}
\widehat{\mathbf{x}}_{k|k}^{b}-\mathbf{x}_{k}=\left( \mathbf{W}%
_{k}^{b}\right) ^{H}\mathbf{v}_{k}-\left( \mathbf{I}-\left( \mathbf{W}%
_{k}^{b}\right) ^{H}\mathbf{H}_{k}\right) \mathbf{w}_{k-1} \\
+\left( \mathbf{I}-\left( \mathbf{W}_{k}^{b}\right) ^{H}\mathbf{H}%
_{k}\right) \mathbf{F}_{k-1}\left( \left( \overline{\mathbf{W}}%
_{k-1}^{b}\right) ^{H}\overline{\mathbf{n}}_{k-1}-\mathbf{G}_{k-1}\overline{%
\mathbf{w}}_{k-2}\right) ,
\end{multline*}%
that is:%
\begin{multline*}
\widehat{\mathbf{x}}_{k|k}^{b}-\mathbf{x}_{k}=\left( \mathbf{W}%
_{k}^{b}\right) ^{H}\mathbf{v}_{k}-\left( \mathbf{I}-\left( \mathbf{W}%
_{k}^{b}\right) ^{H}\mathbf{H}_{k}\right) \mathbf{G}_{k}\overline{\mathbf{w}}%
_{k-1} \\
+\left( \mathbf{I}-\left( \mathbf{W}_{k}^{b}\right) ^{H}\mathbf{H}%
_{k}\right) \mathbf{F}_{k-1}\left( \overline{\mathbf{W}}_{k-1}^{b}\right)
^{H}\overline{\mathbf{n}}_{k-1},
\end{multline*}%
since $\mathbf{G}_{k}\overline{\mathbf{w}}_{k-1}=\mathbf{F}_{k-1}\mathbf{G}%
_{k-1}\overline{\mathbf{w}}_{k-2}+\mathbf{w}_{k-1}$. Moreover, whatever the
constraint $\mathcal{C}_{k}^{1}$ or $\mathcal{C}_{k}^{3}$ considered,
according to (\ref{Dk-1 / Wk-1}):%
\begin{equation*}
\left( \overline{\mathbf{D}}_{k-1}^{b}\right) ^{H}=\left( \mathbf{I}-\left(
\mathbf{W}_{k}^{b}\right) ^{H}\mathbf{H}_{k}\right) \mathbf{F}_{k-1}\left(
\overline{\mathbf{W}}_{k-1}^{b}\right) ^{H}.
\end{equation*}%
Thus:
\begin{multline*}
\widehat{\mathbf{x}}_{k|k}^{b}-\mathbf{x}_{k}=\left( \overline{\mathbf{D}}%
_{k-1}^{b}\right) ^{H}\overline{\mathbf{n}}_{k-1}+\left( \mathbf{W}%
_{k}^{b}\right) ^{H}\left( \mathbf{H}_{k}\mathbf{G}_{k}\overline{\mathbf{w}}%
_{k-1}+\mathbf{v}_{k}\right) \\
-\mathbf{G}_{k}\overline{\mathbf{w}}_{k-1},
\end{multline*}%
that is:%
\begin{multline}
\widehat{\mathbf{x}}_{k|k}^{b}-\mathbf{x}_{k}=\left( \overline{\mathbf{D}}%
_{k-1}^{b}\right) ^{H}\overline{\mathbf{n}}_{k-1}+\left( \mathbf{W}%
_{k}^{b}\right) ^{H}\mathbf{n}_{k}-\mathbf{G}_{k}\overline{\mathbf{w}}_{k-1}
\label{distortionless property xk|k} \\
=\left( \overline{\mathbf{W}}_{k}^{b}\right) ^{H}\overline{\mathbf{n}}_{k}-%
\mathbf{G}_{k}\overline{\mathbf{w}}_{k-1},
\end{multline}%
since $\mathbf{n}_{k}=\mathbf{H}_{k}\mathbf{G}_{k}\overline{\mathbf{w}}%
_{k-1}+\mathbf{v}_{k}$, which proves the distortionless property propagation.

\subsection{From LCKF to linearly constrained LMVDRF}

As mentioned in the introduction, the Fisher initialization (\ref{WLSE x1})
yields the stochastic LMVDRF, which shares the same recursion as the KF,
except at time $k=1$, as recalled in Subsection \ref{S: LMVDRF}. Although
the LMVDRF is sub-optimal in terms of MSE, it has a number of merits \cite%
{Chaumette - TAC}: a) it does not depend on the prior knowledge (first and
second order statistics) on the initial state, b) it may outperform the KF
in case of misspecification of the prior knowledge on $\mathbf{x}_{0}$ \cite[%
Section VI]{Chaumette - TAC}. In other words, the LMVDRF can be pre-computed
and its behaviour can be assessed in advance independently of the prior
knowledge on $\mathbf{x}_{0}$. Interestingly enough, since the
predictor/corrector recursion (\ref{recursion C1 - beg}) propagates the
distortionless property (\ref{distortionless property xk|k}), these results
are still valid regarding the LCKF, which can be looked upon as a linearly
constrained "initial state first and second order statistics" matched
filter. Indeed one can transform a LCKF into a linearly constrained LMVDRF
(LCLMVDRF) provided that $\mathbf{H}_{1}$ and $\mathbf{C}_{\mathbf{v}_{1}}$
are full rank, and $N_{1}$ is large enough to incorporate the distortionless
constraints if the LCKF already verifies some linear constraints at time $k=1
$: $\mathbf{W}_{1}^{H}\mathbf{\Delta }_{1}=\mathbf{T}_{1}$, $\mathbf{\Delta }%
_{1}$ full rank $\rightarrow $ $\mathbf{W}_{1}^{H}\left[ \mathbf{\Delta }%
_{1}~\mathbf{H}_{1}\right] =\left[ \mathbf{T}_{1}~\mathbf{I}\right] $, $%
\left[ \mathbf{\Delta }_{1}~\mathbf{H}_{1}\right] $ full rank. These
features are quite interesting for filtering performance analysis and design
of a LDSS system since they allow to synthesize a wide variety of linearly
constrained infinite impulse response (IIR) distortionless filters which
performance is robust to an unknown initial state.

\section{Deterministic parameters estimation\label{S: Deterministic
parameters estimation}}

If for LDSS models the focus has always been on the LLMSF, in deterministic
parameters estimation, the maximum likelihood estimator (MLE) is the most
used because of its nearly optimal properties in the asymptotic regime \cite%
{OVSN93}\cite{Ren06}. However the MLE suffers from a large computational
cost as it generally requires solving a non-linear multidimensional
optimization problem, which has led to the development of various
sub-optimal techniques to reduce the computational burden \cite{Van Trees
Part IV}. For instance, in the fields of radar, sonar, and wireless
communication, it is common place to design a LMVDRE for the most studied
estimation problem: that of separating the components of data formed from a
linear superposition of individual signals to noisy data \cite{Van Trees
Part IV}\cite{Schreier - Scharf}. This is the reason why, sometimes, the
LMVDRE is also called a deconvolution filter \cite[\S 6]{Van Trees Part IV}%
\cite[\S 5.6]{Schreier - Scharf}. In the case of LDSS models (\ref{DLS -
recursion - xk}-b), the stochastic filtering problem turns into a
deterministic estimation problem if $\mathbf{x}_{k}=\mathbf{x}_{k-1}=\ldots =%
\mathbf{x}_{1}$, where $\mathbf{x}_{1}$ is deterministic and unknown (i.e. $%
\mathbf{w}_{k}=\mathbf{0}$, $k\geq 0$). In this instance, the assumptions (%
\ref{LLMSF - recursive pred/cor cond}) reduces to $\mathbf{C}_{\mathbf{v}%
_{k},\overline{\mathbf{y}}_{k-1}}=\mathbf{C}_{\mathbf{v}_{k},\overline{%
\mathbf{v}}_{k-1}}=\mathbf{0}$,\textbf{\ }$k\geq 2$, which means that the
measurement noise sequence is temporally uncorrelated: $\mathbf{C}_{\mathbf{v%
}_{l},\mathbf{v}_{k}}=\mathbf{C}_{\mathbf{v}_{k}}\delta _{k}^{l}$. Thus, as
already noticed in \cite{Chaumette - TAC}, under temporally uncorrelated
measurement noise and provided that $\mathbf{H}_{1}$ and $\mathbf{C}_{%
\mathbf{v}_{1}}$ are full rank \cite{Chaumette - SPL}, the LMVDRE:
\begin{equation}
\widehat{\mathbf{x}}_{1|k}^{b}=\left( \overline{\mathbf{W}}_{k}^{b}\right)
^{H}\overline{\mathbf{y}}_{k},~\overline{\mathbf{W}}_{k}^{b}=\mathbf{C}_{%
\overline{\mathbf{v}}_{k}}^{-1}\overline{\mathbf{A}}_{k}\left( \overline{%
\mathbf{A}}_{k}^{H}\mathbf{C}_{\overline{\mathbf{v}}_{k}}^{-1}\overline{%
\mathbf{A}}_{k}\right) ^{-1},  \label{x1 - LMVDRE}
\end{equation}%
is the special case of the LMVDRF for which the predictor/corrector
recursion is of the form:
\begin{subequations}
\begin{align}
\widehat{\mathbf{x}}_{1|k}^{b}& =\widehat{\mathbf{x}}_{1|k-1}^{b}+\left(
\mathbf{W}_{k}^{b}\right) ^{H}\left( \mathbf{y}_{k}-\mathbf{H}_{k}\widehat{%
\mathbf{x}}_{1|k-1}^{b}\right)  \\
\mathbf{S}_{1|k}& =\mathbf{H}_{k}\mathbf{P}_{1|k-1}^{b}\mathbf{H}_{k}^{H}+%
\mathbf{C}_{\mathbf{v}_{k}},  \label{WLSE - beg} \\
\mathbf{W}_{k}^{b}& =\mathbf{S}_{1|k}^{-1}\mathbf{H}_{k}\mathbf{P}%
_{1|k-1}^{b}, \\
\mathbf{P}_{1|k}^{b}& =\left( \mathbf{I}-\left( \mathbf{W}_{k}^{b}\right)
^{H}\mathbf{H}_{k}\right) \mathbf{P}_{1|k-1}^{b}.  \label{WLSE - end}
\end{align}%
However, in deterministic parameters estimation, it is well known that the
performance achievable by the LMVDRE \cite[\S\ 6.7]{Van Trees Part IV}
strongly depends on the accurate knowledge on the parametric models of the
measurement equations (\ref{DLS - recursion - yk}), that is on $\mathbf{H}%
_{k}$ and $\mathbf{C}_{\mathbf{v}_{k}}$. For instance, in array processing,
the performance of MVDR beamformers are not particularly robust in the
presence of various types of differences between the model and the actual
environment (array perturbation, direction of arrival mismatch, inaccurate
estimation of $\mathbf{C}_{\mathbf{v}_{k}}$, ...) \cite[\S\ 6.6]{Van Trees
Part IV}. Thus LCMV beamformers have been developed in which additional
linear constraints are imposed to make the MVDR beamformer more robust \cite[%
\S\ 6.7]{Van Trees Part IV}\cite{Vorobyov}. Interestingly enough, the
existence of recursive LCKFs, and more specifically of recursive LCLMVDRFs,
also proves the existence of recursive LCMVEs under temporally uncorrelated
measurement noise, which are obtained by adding at each (or at some)
recursion a set of linear constraints:
\end{subequations}
\begin{multline*}
\widehat{\mathbf{x}}_{1|k}^{b}=\widehat{\mathbf{x}}_{1|k-1}^{b}+\left(
\mathbf{W}_{k}^{b}\right) ^{H}\left( \mathbf{y}_{k}-\mathbf{A}_{k}\widehat{%
\mathbf{x}}_{1|k-1}^{b}\right) \text{ } \\
\text{s.t. }\left( \mathbf{W}_{k}^{b}\right) ^{H}\mathbf{\Delta }_{k}=%
\mathbf{T}_{k},
\end{multline*}%
provided that $\mathbf{\Delta }_{k}$ is full rank and $\mathbf{W}_{k}^{b}$
is computed as follows (\ref{recursion C1 - beg}-\ref{recursion C1 - end}):
\begin{subequations}
\begin{align}
\mathbf{S}_{1|k}& =\mathbf{H}_{k}\mathbf{P}_{1|k-1}^{b}\mathbf{H}_{k}^{H}+%
\mathbf{C}_{\mathbf{v}_{k}},~\mathbb{W}_{k}=\mathbf{S}_{1|k}^{-1}\mathbf{H}%
_{k}\mathbf{P}_{1|k-1}^{b}, \\
\mathbf{W}_{k}^{b}& =\mathbb{W}_{k}+\mathbf{S}_{1|k}^{-1}\mathbf{\Delta }%
_{k}\left( \mathbf{\Delta }_{k}^{H}\mathbf{S}_{1|k}^{-1}\mathbf{\Delta }%
_{k}\right) ^{-1}\left( \mathbf{T}_{k}-\mathbb{W}_{k}^{H}\mathbf{\Delta }%
_{k}\right) ^{H},
\end{align}%
\begin{multline}
\mathbf{P}_{1|k}^{b}=\left( \mathbf{I}-\mathbb{W}_{k}^{H}\mathbf{H}%
_{k}\right) \mathbf{P}_{1|k-1}^{b}+ \\
\left( \mathbf{T}_{k}-\mathbb{W}_{k}^{H}\mathbf{\Delta }_{k}\right) \left(
\mathbf{\Delta }_{k}^{H}\mathbf{S}_{1|k}^{-1}\mathbf{\Delta }_{k}\right)
^{-1}\left( \mathbf{T}_{k}-\mathbb{W}_{k}^{H}\mathbf{\Delta }_{k}\right)
^{H}.
\end{multline}%
However, the disadvantage of using multiple linear constraints is that
additional degrees of freedom are used by the LCMVE or the LCKF in order to
satisfy these constraints, which increases the minimum MSE achieved.

\subsection{The deterministic least-squares problem}

For sake of completeness, let us recall that, under temporally uncorrelated
measurement noise, the LMVDRE (\ref{x1 - LMVDRE}) and the WLSE:
\end{subequations}
\begin{equation*}
\widehat{\mathbf{x}}_{1|k}^{b}=\arg \underset{\mathbf{x}_{1}}{\min }\left\{
\tsum\limits_{l=1}^{k}\left( \mathbf{y}_{l}-\mathbf{H}_{l}\mathbf{x}%
_{1}\right) ^{H}\mathbf{C}_{\mathbf{v}_{l}}^{-1}\left( \mathbf{y}_{l}-%
\mathbf{H}_{l}\mathbf{x}_{1}\right) \right\} ,
\end{equation*}%
are identical (duality) \cite[\S 3.4]{Kailath - Sayed - Hassibi}\cite[\S 4]%
{Gibbs}. As a consequence of this, provided that $\mathbf{H}_{1}$ and $%
\mathbf{C}_{\mathbf{v}_{1}}$ are full rank, the WLSE and the regularized
WLSE (RWLSE) are primarily special cases of the LMVDRF \cite{Chaumette - TAC}%
, and their relation to the KF highlighted in \cite{Sayed - Kailath} is
actually purely formal. The extension of this result to the RWLSE \cite[\S %
2.4]{Kailath - Sayed - Hassibi}\cite{Sayed - Kailath}:
\begin{equation*}
\widehat{\mathbf{x}}_{1|k}^{b}=\arg \underset{\mathbf{x}_{1}}{\min }\left\{
\begin{array}{l}
\left( \mathbf{c}-\mathbf{x}_{1}\right) ^{H}\mathbf{\Sigma }^{-1}\left(
\mathbf{c}-\mathbf{x}_{1}\right) + \\
\tsum\limits_{l=1}^{k}\left( \mathbf{y}_{l}-\mathbf{H}_{l}\mathbf{x}%
_{1}\right) ^{H}\mathbf{C}_{\mathbf{v}_{l}}^{-1}\left( \mathbf{y}_{l}-%
\mathbf{H}_{l}\mathbf{x}_{1}\right)%
\end{array}%
\right\}
\end{equation*}%
where $\mathbf{\Sigma }$ is an Hermitian invertible matrix, is simply
obtained by adding a fictitious observation at time $k=0$: $\mathbf{y}_{0}=%
\mathbf{H}_{0}\mathbf{x}_{1}+\mathbf{v}_{0}$, $\mathbf{C}_{\mathbf{v}_{0}}=%
\mathbf{\Sigma }$, $\mathbf{y}_{0}=\mathbf{c}$, $\mathbf{H}_{0}=\mathbf{I}$,
and by starting the recursion at time $k=0$: $\widehat{\mathbf{x}}_{1|0}^{b}=%
\mathbf{P}_{1|0}^{b}\mathbf{H}_{0}^{H}\mathbf{C}_{\mathbf{v}_{0}}^{-1}%
\mathbf{y}_{0}=\mathbf{c}$, $\mathbf{P}_{1|0}^{b}=\left( \mathbf{H}_{0}^{H}%
\mathbf{C}_{\mathbf{v}_{0}}^{-1}\mathbf{H}_{0}\right) ^{-1}=\mathbf{\Sigma }$
\cite{Chaumette - TAC}.

\section{An illustrative example}

In the case of LDSS model (\ref{DLS - recursion - xk}-b), turning the KF
into the LMVDRF thanks to the Fisher initialization (\ref{WLSE x1}), can be
regarded as a first step towards the robustification of KF, namely to an
unknown initial state. To some extent, the LCKF can also robustify the KF in
the presence of parametric modelling errors in system matrices: $\mathbf{F}%
_{k}\triangleq \mathbf{F}_{k}\left( \mathbf{\omega }\right) =\left[ \mathbf{f%
}_{k}^{1}\left( \mathbf{\omega }\right) ~\ldots ~\mathbf{f}%
_{k}^{P_{k}}\left( \mathbf{\omega }\right) \right] $ and $\mathbf{H}%
_{k}\triangleq \mathbf{H}_{k}\left( \mathbf{\theta }\right) =\left[ \mathbf{h%
}_{k}^{1}\left( \mathbf{\theta }\right) ~\ldots ~\mathbf{h}%
_{k}^{P_{k}}\left( \mathbf{\theta }\right) \right] $, where $\mathbf{\omega }
$ and $\mathbf{\theta }$ are supposed to be deterministic vector values
determined via an ad hoc calibration process. In many cases, such
calibration process provides estimates $\widehat{\mathbf{\omega }}=\mathbf{%
\omega }+d\widehat{\mathbf{\omega }}$ and $\widehat{\mathbf{\theta }}=%
\mathbf{\theta }+d\widehat{\mathbf{\theta }}$ of the true values $\mathbf{%
\omega }$ and $\mathbf{\theta }$, which means that the predictor/corrector
recursion (\ref{recursion C1 - beg}) is updated according to $\mathbf{F}%
_{k-1}\left( \widehat{\mathbf{\omega }}\right) $ and $\mathbf{H}_{k}\left(
\widehat{\mathbf{\theta }}\right) $, i.e. $\mathbf{W}_{k}^{b}\triangleq
\mathbf{W}_{k}^{b}\left( \widehat{\mathbf{\omega }},\widehat{\mathbf{\theta }%
}\right) $. If the calibration process is accurate enough, i.e. $d\widehat{%
\mathbf{\omega }}$ and $d\widehat{\mathbf{\theta }}$ are small, then the
true state and measurement matrices $\mathbf{F}_{k}\left( \mathbf{\omega }%
\right) $ and $\mathbf{H}_{k}\left( \mathbf{\theta }\right) $ differ from
the assumed ones via first order Taylor series. In this circumstance, the
following additional linear constraints:
\begin{equation*}
\mathbf{W}_{k}^{H}\left[ \frac{\partial \mathbf{h}_{k}^{1}\left( \widehat{%
\mathbf{\theta }}\right) }{\partial \mathbf{\theta }}~\ldots ~\frac{\partial
\mathbf{h}_{k}^{P_{k}}\left( \widehat{\mathbf{\theta }}\right) }{\partial
\mathbf{\theta }}\right] =\mathbf{0},
\end{equation*}%
\begin{subequations}
\begin{equation*}
\mathbf{W}_{k}^{H}\mathbf{H}_{k}\left( \widehat{\mathbf{\theta }}\right) %
\left[ \frac{\partial \mathbf{f}_{k-1}^{1}\left( \widehat{\mathbf{\omega }}%
\right) }{\partial \mathbf{\omega }}~\ldots ~\frac{\partial \mathbf{f}%
_{k-1}^{P_{k-1}}\left( \widehat{\mathbf{\omega }}\right) }{\partial \mathbf{%
\omega }}\right] =\mathbf{0},
\end{equation*}%
yields:
\end{subequations}
\begin{equation*}
\mathbf{W}_{k}^{H}\mathbf{y}_{k}\simeq \mathbf{W}_{k}^{H}\left( \mathbf{H}%
_{k}\left( \widehat{\mathbf{\theta }}\right) \left( \mathbf{F}_{k-1}\left(
\widehat{\mathbf{\omega }}\right) \mathbf{x}_{k-1}+\mathbf{w}_{k-1}\right) +%
\mathbf{v}_{k}\right) ,
\end{equation*}%
which means that the predictor/corrector recursion (\ref{recursion C1 - beg}%
) has become robust to (small) parametric modelling error on the state and
measurement matrices. Note that the proposed approach encompasses the LDSS
model introduced in \cite{Theodor - Shaked}\cite{Xie94}:
\begin{subequations}
\begin{eqnarray*}
\mathbf{x}_{k} &=&\left( \mathbf{F}_{k}+\Delta \mathbf{F}_{k}\right) \mathbf{%
x}_{k-1}+\mathbf{w}_{k-1} \\
\mathbf{y}_{k} &=&\left( \mathbf{H}_{k}+\Delta \mathbf{H}_{k}\right) \mathbf{%
x}_{k}+\mathbf{v}_{k}
\end{eqnarray*}%
where $\mathbf{x}_{k}\in
\mathbb{R}
^{P}$, $\mathbf{y}_{k}\in
\mathbb{R}
^{N}$. The matrices $\Delta \mathbf{F}_{k}$ and $\Delta \mathbf{H}_{k}$
represent the parameter uncertainties and have the following structure \cite%
{Theodor - Shaked}: $\Delta \mathbf{F}_{k}=\mathbf{A}_{1,k}\mathbf{B}_{k}%
\mathbf{C}_{k}$ and $\Delta \mathbf{H}_{k}=\mathbf{A}_{2,k}\mathbf{B}_{k}%
\mathbf{C}_{k}$, where $\mathbf{A}_{1,k}$, $\mathbf{A}_{2,k}$ and $\mathbf{C}%
_{k}$ are known matrices of the appropriate dimensions, and $\mathbf{B}_{k}$
are unknown matrices satisfying $\mathbf{B}_{k}\mathbf{B}_{k}^{T}\leq
\mathbf{I}$. Then the LCKF may provide an alternative design of a linear
filter such that the variance of the filtering error is guaranteed to be
within a certain bound for all admissible uncertainties. Indeed, provided
that:
\end{subequations}
\begin{equation*}
\mathbf{W}_{k}^{H}\left[ \mathbf{A}_{2,k}~\mathbf{H}_{k}\mathbf{A}_{1,k}%
\right] =\mathbf{0},
\end{equation*}%
has a non trivial solution, that is $N>rank\left( \left[ \mathbf{A}_{2,k}~%
\mathbf{H}_{k}\mathbf{A}_{1,k}\right] \right) $ and $\left[ \mathbf{A}_{2,k}~%
\mathbf{H}_{k}\mathbf{A}_{1,k}\right] $ is full rank, then:
\begin{equation*}
\mathbf{W}_{k}^{H}\mathbf{y}_{k}=\mathbf{W}_{k}^{H}\left( \mathbf{H}%
_{k}\left( \mathbf{F}_{k}\mathbf{x}_{k-1}+\mathbf{w}_{k-1}\right) +\mathbf{v}%
_{k}\right)
\end{equation*}%
and the LCKF does not depends on $\Delta \mathbf{F}_{k}$ and $\Delta \mathbf{%
H}_{k}$ any longer.

\section{Conclusion}

We introduced the general form of the LCKF for LDSS models. Since the LCMVE
is a special case of the LCKF, the use of LCKF, among other things, opens
access to the abundant literature on LCMVE in the deterministic framework
which can be transposed to the stochastic framework in order to provide
alternative solutions to $H_{\infty }$ filter and UFIR filter to robustify
the KF.

\end{document}